\documentclass[reprint,nofootinbib,superscriptaddress,amsmath,amssymb,aps,prd]{revtex4-1}
\usepackage{amsmath,mathrsfs,amsbsy,color,graphicx,bm,amsthm,amsfonts}
\usepackage{bbm}
\usepackage{times}
\usepackage{units}
\usepackage{dcolumn}
\usepackage{graphicx}
\usepackage{epsfig}
\usepackage{epstopdf}
\usepackage[colorlinks,linkcolor=red,anchorcolor=green,citecolor=blue,CJKbookmarks=True]{hyperref}
\DeclareMathSymbol{\shortminus}{\mathbin}{AMSa}{"39}
\usepackage{mathrsfs}
\usepackage{braket}
\usepackage{amssymb}
\usepackage{txfonts}
\usepackage{float}
\usepackage{enumitem}
\usepackage{multirow}

\newcommand{\meq}[1]{(\ref{#1})}

\newcommand{\non}{\nonumber \\}

\newcommand{\hsp}{\hspace{0.1mm}}

\newcommand{\pp}{\partial}

\begin{document}

\title{Lorentz violation alleviates gravitationally induced entanglement degradation}

\author{Wentao Liu}
\affiliation{Department of Physics, Key Laboratory of Low Dimensional Quantum Structures and Quantum Control of Ministry of Education, and Synergetic Innovation Center for Quantum Effects and Applications, Hunan Normal
University, Changsha, Hunan 410081, P. R. China}

\author{Cuihong Wen}
\affiliation{Department of Physics, Key Laboratory of Low Dimensional Quantum Structures and Quantum Control of Ministry of Education, and Synergetic Innovation Center for Quantum Effects and Applications, Hunan Normal
University, Changsha, Hunan 410081, P. R. China}


\author{Jieci Wang}
\email{jcwang@hunnu.edu.cn}
\affiliation{Department of Physics, Key Laboratory of Low Dimensional Quantum Structures and Quantum Control of Ministry of Education, and Synergetic Innovation Center for Quantum Effects and Applications, Hunan Normal
University, Changsha, Hunan 410081, P. R. China}


\begin{abstract}

Lorentz violation is a significant phenomenon in the framework of quantum physics, with implications for fundamental symmetries. 
In this paper, we explore the effects of Lorentz violation on quantum entanglement through a black hole spacetime that is coupled with a Lorentz-violating field.
We establish the relationship between the Hartle-Hawking vacuum state and the Boulware number states for this case, and employ the near horizon approximation in an appropriate form to rewrite the black hole metric into a Rindler-like form.
Subsequently, using this revised metric, the analytical forms of logarithmic negativity and mutual information are derived and plotted as functions of Rob's distance from the $ r=0 $ point.
Based on the results, we find that the coupling between spacetime and the Lorentz-violating vector field alleviates gravity-induced entanglement degradation. 
At high mode frequencies, the effects of Lorentz violation are negligible, but they become significant at low frequencies. 
This suggests that investigating Lorentz violation at astrophysical scales requires low-frequency detectors, as the low energy of these fields enhances the significance of the Lorentz-violating field's non-zero vacuum expectation value.


\end{abstract}

\vspace*{0.5cm}

 \maketitle

\section{INTRODUCTION}


%

Quantum information theory explores how the principles of quantum mechanics, such as superposition and entanglement, can be applied to information processing and transmission, enabling the encoding and manipulation of information through quantum states \cite{Boschi:1997dg,Pan:2001nrz,Hu:2018mni}.
As our understanding deepened, the need to incorporate relativistic concepts emerged, leading to the development of relativistic quantum information theory \cite{Fuentes-Schuller:2004iaz}. 
A key area of research in this field is the study of quantum correlations from a non-inertial perspective, offering new insights into the intersection of relativity and quantum mechanics \cite{Alsing:2006cj,Wang:2009qg,Martin-Martinez:2009hfq,Martin-Martinez:2010yva,Esfahani:2010gt,Wang:2009wg,Esfahani:2010gt,He:2016ujs,Wang:2017tfv,Belfiglio:2021mnr,Bueley:2022ple,Wu:2023sye,Wu:2023spa,Mondal:2022aev,Wu:2022lmc,Babakan:2024abb,Wu:2024qhd,Wu:2024BF,Liu:2024pse}.

In various studies, one topic that has attracted attention is the entanglement degradation phenomenon in a bipartite system where one observer is uniformly accelerated; this phenomenon, sometimes called Unruh decoherence, is strongly related to the Unruh effect \cite{Crispino:2007eb,Kollas:2022wgj}.
The goal was to shift the system's state into Rindler spacetime to better understand how acceleration effect influences quantum entanglement.
Rather than focusing on spin entanglement, these studies examined entanglement between number states, using Fock states to capture the quantum correlations under acceleration \cite{Fuentes-Schuller:2004iaz,Alsing:2006cj,Wang:2009qg}.
Interestingly, the behavior of an observer approaching the event horizon in a black hole spacetime is analogous to the infinite acceleration limit in Rindler spacetime.
The system consists of two observers, Alice and Rob, who share a maximally entangled state in the asymptotically flat region. 
Alice remains in this region, while Rob moves to a fixed radial distance outside the event horizon of the black hole, leading to entanglement degradation due to the black hole's gravitational influence.

On the other hand, understanding the modifications to gravitational theory by introducing quantum properties is essential in fundamental physics \cite{Kostelecky19891}.
One significant area of study is the breaking of Lorentz symmetry, which sheds light on the underlying mechanisms of gravity \cite{Kostelecky2001,Choi:2015bga,DelPorro:2023lbv,Tian:2022gfa,Agarwalla:2023wft,Jiang:2024agx}. 
A key model in this field is the bumblebee gravity theory \cite{Kostelecky:2003fs,Bluhm:2004ep}, which introduces a nonminimally coupled vector field that can spontaneously break Lorentz symmetry by acquiring a nonzero vacuum expectation value (VEV). 
This results in an anisotropic energy-momentum tensor and extends the framework of general relativity with new interactions. 
The exploration of exact solutions in Bumblebee gravity, including both spherically symmetric and rotating cases \cite{Casana2018,Ovgun2019,Gullu2020,Poulis:2021nqh,Maluf2021,Xu:2022frb,Ding2022}, has driven significant advances in black hole physics, making Lorentz symmetry breaking a vibrant area of ongoing research \cite{Liu:2022dcn,Mai:2023ggs,Xu:2023xqh,Zhang:2023wwk,Lin:2023foj,Chen:2023cjd,Chen2020,Wang:2021gtd,Liu:2024oeq,Mai:2024lgk,Liang:2022gdk}.

Given that black holes serve as focal points where gravitational and other effects intersect, and are often used as testbeds for corresponding theories, a natural question arises: how will quantum modifications of black holes affect entanglement degradation near the event horizon?
To achieve this,  we will compute the quantum entanglement and mutual information as a precise function of four physical parameters: the distance of Rob from the event horizon, the mass of the black hole, the Lorentz-violating parameter, and the mode frequency that characterizes the entanglement between Rob's state and Alice's field state. 
The Lorentz-violating parameter includes the non-zero VEV from the bumblebee field and the non-minimal coupling constant \cite{Casana2018}.
As a result of this study, we will derive the explicit form of quantum correlations as a function of the aforementioned physical parameters. 

The organization of the paper is as follows.
In Sec. \ref{sec2}, we briefly introduced the Einstein-Bumblebee gravity model, which is a modification of general relativity (GR) by a Lorentz-violating vector field.
In Sec. \ref{sec3}, we reformulate the Einstein-Bumblebee black hole in Rindler coordinates and explore the corresponding relationships among different vacuum states.
In Sec. \ref{sec4}, we present the analytical expressions for the logarithmic negativity and mutual information of scalar and Dirac fields in the Einstein-Bumblebee black hole background, and plot these quantities as functions of the observer's distance from $ r=0 $.
Finally, our conclusions and outlooks are given in Sec. \ref{sec5}.

\section{Black hole spacetime}\label{sec2}
Here we briefly review the Einstein-Bumblebee gravity model. 
This model is known as an example that extends the standard formalism of GR. 
The action for the bumblebee field $B_\mu$ coupled to gravity can be described as \cite{Casana2018}
\begin{equation}
\begin{aligned}\label{Action}
\mathcal{S}_B=&\int d^4x \sqrt{-g}\left[\frac{1}{2\kappaup}\left(R-2\Lambda\right)+\frac{\varrho}{2\kappa} B^\mu B^\nu R_{\mu\nu} \right.\\
&~~~\left. -\frac{1}{4}B^{\mu\nu}B_{\mu\nu}-V\left(B^\mu B_\mu\pm b^2\right)\right],
\end{aligned}
\end{equation}
where $ \kappaup=8\pi G_N $ is the gravitational coupling constant that can be set to $ G_N =1$ without loss of generality.
$ \Lambda $ is the cosmological constant and $ \varrho $ is the real coupling constant which controls the non-minimal gravity interaction to the bumblebee field $B_\mu$. 

It is worth noting that the potential $ V $, designed to ensure a non-zero VEV for the bumblebee field $\langle B_\mu\rangle\equiv b_\mu$, reaches its minimum at $B_\mu B^\mu= \mp b^2$, where $b$ is a positive real constant and the $ \pm $ sign implies that $ B_\mu $ is timelike or spacelike, respectively. 
This nonzero VEV introduces a preferred direction in spacetime, thereby spontaneously breaking Lorentz symmetry in the Bumblebee model. 
Such spontaneous Lorentz violation is conceptually distinct from breaking the gauge invariance of $ B_{\mu} $: gauge transformations act on the field itself $ (B_{\mu}\longrightarrow B_{\mu}+\pp_\mu \chi) $, while Lorentz transformations act on the spacetime coordinates. 
Thus, even though the bumblebee action can exhibit a form of gauge violation, the key point here is that the nonzero VEV of $ B_{\mu} $ picks out a specific frame, explicitly violating Lorentz invariance. For further discussions on spontaneous Lorentz breaking in Bumblebee models, we refer to Ref. \cite{Kostelecky:2003fs}.

Toking of the variational $g_{\mu\nu}$ and $B_\mu$ yields the gravitational equation and bumblebee field equation:
\begin{align}
& R_{\mu\nu}-\frac{1}{2}g_{\mu\nu}\left(R-2\Lambda\right)=\kappa T^{B}_{\mu\nu} \label{EinsteinEQ}, \\ \label{VEq}
& \nabla^\mu B_{\mu\nu}=2V'B_\nu-\frac{\varrho}{\kappa}B^\mu R_{\mu\nu}.
\end{align}
 $T^{B}_{\mu\nu}$ is the bumblebee energy momentum tensor, which have the following form:
\begin{equation}
\begin{aligned}\label{TBab}
T^{B}_{\mu\nu}=
& B_{\mu\alpha}B^\alpha\hsp_\nu-\frac{1}{4}g_{\mu\nu}B^{\alpha\beta}B_{\alpha\beta}-g_{\mu\nu}V+2B_\mu B_\nu V' \\
&+\frac{\varrho}{\kappa}
\left[\frac{1}{2}g_{\mu\nu}B^{\alpha}B^{\beta}R_{\alpha\beta}-B_\mu B^\alpha R_{\alpha \nu}-B_\nu B^\alpha R_{\alpha \mu}\right.\\
& \left.+\frac{1}{2}\nabla_\alpha\nabla_\mu\left(B^\alpha B_\nu\right)
+\frac{1}{2}\nabla_\alpha\nabla_\nu\left(B^\alpha B_\mu\right)\right.\\
&\left.-\frac{1}{2}\nabla^2\left(B_\mu B_\nu\right)-\frac{1}{2}g_{\mu\nu}\nabla_\alpha\nabla_\beta\left(B^\alpha B^\beta\right)  \right].
\end{aligned}
\end{equation}
An exact spherically symmetric black hole solution has been constructed by Casana et al. \cite{Casana2018}, and it is referred to as the Schwarzschild-like black holes.
The radial bumblebee field $ B_\mu $ can be written as
\begin{align}
B_\mu=b_\mu=\left\{0,b\sqrt{\frac{(1+\ell)}{F(r)}},0,0\right\},
\end{align}
and the metric is given 
\begin{align}\label{ds2MOG}
g_{\mu\nu}=&\text{diag}\left\{-F(r),\frac{(1+\ell)}{F(r)},r^2,r^2 \sin^2\theta \right\},
\end{align}
where metric function is $ F(r)=1-2M/r $ and the symbol $ \ell=\varrho b^2 $ represents the Lorentz violation parameter.
From a gravitational perspective, this solution is sustained by an anisotropic energy-momentum tensor, which reflects the manifestation of the bumblebee field in the spacetime geometry. 
In particular, $ \ell=\varrho b^2 $ quantifies the degree of Lorentz violation, while the radial profile $ b_{\mu} $ enforces an anisotropic energy-momentum distribution. 
As a result, the vacuum state no longer maintains the usual isotropy and homogeneity under Lorentz transformations.

Now we consider the surface gravity of black holes, which is generally regarded as the acceleration at the event horizon.
The four-velocity of a stationary particle is
\begin{align}
u^\mu=\left\{u^0,0,0,0\right\},
\end{align}
where  $ u^0 $ is determined by the normalized relation $ u^\mu u_\mu=-1 $.
The corresponding four-acceleration is
\begin{align}\label{amu}
a^\nu=u^\mu\nabla_\mu u^\nu=u^\mu\partial_\mu u^\nu+\Gamma^\nu_{\mu\rho}u^\mu u^\rho,
\end{align}
and the surface gravity is given by $ \kappa=\frac{1}{4M(1+\ell)} $.
Note that in an arbitrary theory of gravity, when quantum particle creation effects are considered, the Hawking temperature of a black hole with constant surface gravity is \(T=\kappa/2\pi\), as discussed in \cite{Wald:1993nt}.

Notably, the metric \meq{ds2MOG} can be transformed into that of a black hole with a global monopole by rescaling the mass and radial coordinate. 
Specifically, we introduce the following transformations:
\begin{equation}
r\longrightarrow\frac{1}{\sqrt{1+\ell}}R,~~~~~~\mathcal{M}=\sqrt{1+\ell}M,~~~~~~\eta=\frac{1}{\sqrt{1+\ell}},
\end{equation}
where $ \mathcal{M} $ denotes the redefined mass and $ \eta $ acts as a monopole-like parameter. 
This leads to
\begin{align}
\bar{g}_{\mu\nu}=&\text{diag}\left\{-\bar{F}(R),\frac{1}{\bar{F}(R)},\eta^2R^2,\eta^2R^2 \sin^2\theta \right\},
\end{align}
with $ \bar{F}(R)=1-2\mathcal{M}/R $.
Global monopoles are a type of topological defect that may form in the early universe through the spontaneous breaking of the global $ O(3) $ symmetry down to $ U(1) $ \cite{Kibble:1976sj,Vilenkin:1984ib}. 
Both the bumblebee field and the global monopole arise from spontaneous symmetry breaking, yet involve different symmetry groups. 
Consequently, two distinct physical mechanisms could lead to similar effects on gravity-induced entanglement degradation. 
Thus far, the influence of global monopoles on entanglement degradation remains largely unexplored, and we hope that our findings will provide a valuable reference and inspiration for further research in this direction. 
Indeed, in their study of perturbation dynamics in bumblebee gravity, M. Wang et al. \cite{Zhang:2023wwk} demonstrated that these mechanisms can produce analogous physical outcomes, particularly in the context of QNMs and late-time tails.

\section{THE ``BLACK HOLE LIMIT'': TRANSLATION Rindler-Kruskal}\label{sec3}
In this section, we shall follow the analysis used in \cite{Martin-Martinez:2010yva,Sen:2023sfb}.
By utilizing the construction demonstrated below in Lorentz-violating spacetime, we will be able to calculate the entanglement loss between a free-falling observer and another one placed at a fixed distance from the event horizon as a function of distance, and study the behavior of quantum correlations in the presence of Einstein-Bumblebee black holes \footnote{S. Gangopadhyay et al. provide an interesting construction for spherically symmetric cases that \cite{Sen:2023sfb}, however, is incompatible with the Einstein-Bumblebee black hole, which is more general spherically symmetric cases.}.
The line element for a static Einstein-Bumblebee black hole is given by
\begin{align}\label{ds22}
ds^2=-F(r)dt^2+\frac{(1+\ell)}{F(r)}dr^2+r^2d\Omega^2,
\end{align}
where $d\Omega^2$ represents the angular part of the metric on the unit sphere.
Employing the near horizon approximation, the metric function can be represented in the following form:
\begin{align}
F(r)\simeq(r-r_h)F'(r_h),
\end{align}
where $ r_h=2M $ is the event horizon of the black hole.
Because of the symmetry of the problem, we will restrict the analysis to the radial coordinate; near the horizon, the radial part of the metric is expressed as $ ds^2=-(r-r_h)F'(r_h)dt^2+(1+\ell)(r-r_h)^{-1}F'(r_h)^{-1}dr^2 $.
Then we define a coordinate transformation as follows:
\begin{align}\label{drdz}
d r=\frac{\zeta F'(r_h)}{2(1+\ell)}d\zeta.
\end{align}
By substituting equation \meq{drdz} and its integral form into the radial line element, we can obtain
\begin{equation}
\begin{aligned}
ds^2=&-\left[\frac{c_1-r_h}{F'(r_h)^{-1}}+\frac{\zeta^2 F'(r_h)^2}{4(1+\ell)}\right]dt^2\\
&+\left[1+\frac{4(c_1-r_h)(1+\ell)}{\zeta^2 F'(r_h)} \right]^{-1}d\zeta^2,
\end{aligned}
\end{equation}
where $ c_1 $ is an arbitrary integration constant.
By setting $ c_1=r_h $ and considering surface gravity $ \kappa=\frac{1}{4M(1+\ell)} $, the line elements are simplified to
\begin{align}
ds^2=-(1+\ell)\zeta^2 \kappa^2dt^2+d\zeta^2.
\end{align}
We can choose to write the line elements in terms of the proper time $ \tau $ for an observer at position $ r_0 $ as follows:
\begin{align}\label{dstau}
ds^2=-\frac{(1+\ell)\kappa^2\zeta^2}{F_0}d\tau^2+d\zeta^2,
\end{align}
which corresponds to a Rindler metric with an acceleration parameter $ \sqrt{(1+\ell)} \kappa/\sqrt{F_0} $, where $ d\tau=\sqrt{F_0}dt $ and $ F_0\equiv F(r_0) $.
To elucidate the physical significance of the acceleration parameter, it is essential to calculate the proper acceleration at $ r=r_0 $ for an observer outside a black hole.

For an accelerated observer positioned at arbitrary fixed position $ r $, the value of the proper acceleration is given by
\begin{align}
a=\sqrt{a_\mu a^\mu},
\end{align}
where $ a^\mu $ is defined in Eq.~\meq{amu}.
Evaluating the four acceleration of the observer yields,
\begin{align}\label{admu}
a^\mu=\left\{0,\frac{F'(r)}{2(1+\ell)},0,0 \right\}, && a_\mu=\left\{0,\frac{F'(r)}{2F(r)},0,0 \right\},
\end{align}
and the proper acceleration of the observer is given by
\begin{align}
a(r)=\sqrt{a^\mu a^\nu g_{\mu\nu}}=\frac{F'(r)}{2\sqrt{(1+\ell)F(r)}}.
\end{align}
If the observer at $ r=r_0 $ is close to the event horizon ($ r_0 \approx r_h $), we can derive the following relation:
\begin{align}
F'(r)\simeq \frac{\pp}{\pp r}\left[(r-r_h)F'(r_h)\right]=F'(r_h)=2(1+\ell)\kappa.
\end{align}
Under this approximation, \eqref{dstau} can be rewritten as
\begin{align}
ds^2=-(a_0 \zeta)^2d\tau^2+d\zeta^2,
\end{align}
where the proper acceleration $ a_0 $ for an observer is given by
\begin{align}
a_0\equiv a(r_0)=\frac{\sqrt{(1+\ell)}\kappa}{\sqrt{F_0}}.
\end{align}
Our approach aligns with that described in Ref. \cite{Martin-Martinez:2010yva} and further incorporates the effects of Lorentz violation corrections.
This shows that near the event horizon, the Einstein-Bumblebee metric can be approximated by a Rindler metric, with its acceleration parameter, corrected for Lorentz violation, matching the proper acceleration of an observer at $ r_0 $, close to the event horizon.

Next, to define timelike vectors, we need to shift our focus to the Kruskal framework.
By defining a tortoise coordinate $ r_*=\sqrt{(1+\ell)}\int F(r)^{-1} dr $ and using the light-cone coordinates $ u=t-r_* $ and $ v=t+r_* $ , the radial part of the line element \eqref{ds22} can be rewritten as follows
\begin{align}\label{mFds}
ds^2=-F(r)dudv.
\end{align}
Introducing the generalized light-like Kruskal coordinates
\begin{equation}
\begin{cases}
 &\mathcal{U}=-\frac{1}{\sqrt{(1+\ell)}\kappa}e^{-\sqrt{(1+\ell)}\kappa u}\\
 &\mathcal{V}=\frac{1}{\sqrt{(1+\ell)}\kappa}e^{\sqrt{(1+\ell)}\kappa v}
\end{cases}
\Rightarrow
\begin{cases}
  & du=e^{\sqrt{(1+\ell)}\kappa u}d\mathcal{U}\\
  &dv=e^{-\sqrt{(1+\ell)}\kappa v}d\mathcal{V}
\end{cases},
\end{equation}
near the horizon the Eq.~\meq{mFds} can be expressed as 
\begin{align}\label{ds222}
ds^2=-F(r)e^{-2\sqrt{(1+\ell)}\kappa r_*}d\mathcal{U}d\mathcal{V}\simeq-e^{-1}d\mathcal{U}d\mathcal{V}.
\end{align}
Note that $ e^{-2\sqrt{(1+\ell)}\kappa r_*}=2Me^{-r/(2M)}/(r-2M) $, and the symbol $ \simeq $ indicates the use of a Taylor expansion in the calculation.
Here, following Ref. \cite{Martin-Martinez:2010yva}, we can similarly define the physical timelike vectors for the three regions:
\begin{equation}
\begin{aligned}
\pp_{\hat{t}} &\propto \pp_\mathcal{U} + \pp_\mathcal{V},\\
\pp_t &\propto \left( \mathcal{U}\pp_\mathcal{U} + \mathcal{V}\pp_\mathcal{V} \right),
\end{aligned}
\end{equation}
including $ -\pp_t $ as well. 
Through these different physical timelike vectors, three types of vacuum states can be defined: the Hartle-Hawking vacuum $ \ket{0}_\text{H} $, the Boulware vacuum $ \ket{0}_\text{B} $, and the anti-Boulware vacuum $ \ket{0}_{\bar{\text{B}}} $, respectively.
This is analogous to the vacuum states $ \ket{0}_\text{M} $, $ \ket{0}_\text{I} $, and $ \ket{0}_\text{IV} $ in the Rindler case, and the corresponding relation with the standard Alice-Rob-antiRob notation is as follows\cite{Martin-Martinez:2010yva}:
\begin{equation}
\begin{aligned}\label{ABBt}
&\ket{0}_\text{A}\leftrightarrow\ket{0}_\text{M}\leftrightarrow\ket{0}_\text{H},\\
&\ket{0}_\text{R}\leftrightarrow\ket{0}_\text{I~}\leftrightarrow\ket{0}_\text{B},\\
&\ket{0}_{\bar{\text{R}}}\leftrightarrow\ket{0}_\text{IV}\leftrightarrow\ket{0}_{\bar{\text{B}}}.
\end{aligned}
\end{equation}
The basis transformation between Hartle-Hawking modes and Boulware modes is entirely analogous to that between Minkowskian modes and Rindler modes, characterized by the acceleration parameter $ a_0=\frac{\sqrt{(1+\ell)}\kappa}{\sqrt{F_0}} $.

To begin with, the scalar field satisfies the Klein-Gordon equation. 
At this juncture, the metric form \meq{ds222}, describing the Lorentz-violating spacetime, is consistent with that of the Rindler spacetime. 
Therefore, we can apply the method of calculating the vacuum state and the first excitation of a scalar field in the Rindler case \cite{Fuentes-Schuller:2004iaz} to the Lorentz-violating spacetime geometry, thereby obtaining
\begin{align}\label{sk0}
\ket{0}^{\omega_i}_\text{H}=\frac{1}{\cosh\sigma_{s,i}}\sum_n \tanh^n\sigma_{s,i}\ket{n}^{\omega_i}_\text{B}
\ket{n}^{\omega_i}_{\bar{\text{B}}},
\end{align}
where $ \ket{0}_\text{H}=\otimes_j\ket{0}_\text{H}^{\omega_j} $ and 
\begin{equation}
\begin{aligned}\label{tansi}
\tanh\sigma_{s,i}&=\exp{\left(-\frac{\pi \omega_i}{a_0}\right)}=\exp\left(-\frac{\pi\omega_i\sqrt{F_0}}{\sqrt{(1+\ell)}\kappa}\right)\\
&=\exp{\left[ -4\pi M\omega_i \sqrt{(1+\ell)}\sqrt{1-\frac{2M}{r_0} } \right]}.
\end{aligned}
\end{equation}
The above result includes a Lorentz violation correction and can derive from a direct analogy with the corresponding result in the Minkowski-Rindler scenario.
The process of generating the unprimed one-particle state of Hartle-Hawking, defined within the basis $ \left\{ \psi^\text{H}_{\omega_j}, \psi'^{\text{H}}_{\omega_j} \right\} $, is achieved by the application of the respective creation operator onto the vacuum state. 
Furthermore, this particular state can be converted into the Boulware basis,
\begin{equation}\label{sk1}
\ket{1}_\text{H}^{\omega_i} = \frac{1}{\cosh^2\sigma_{s,i}} \sum_{n=0}^{\infty} \tanh^n\sigma_{s,i} \sqrt{n+1} \ket{n+1}_\text{B}^{\omega_i} \ket{n}^{\omega_i}_{\bar{\text{B}}}.
\end{equation}

For the Dirac field, the vacuum state and one-particle state, similar to those described in Eqs. \eqref{sk0} and \eqref{sk1}, are given by the following formula \cite{Alsing:2006cj}:
\begin{align}
\begin{aligned}
\ket{0}^{\omega_i}_\text{H}=&\cos\sigma_{d,i}\ket{0}^{\omega_i}_\text{B}\ket{0}^{\omega_i}_{\bar{\text{B}}}+\sin\sigma_{d,i}\ket{1}^{\omega_i}_\text{B}\ket{1}^{\omega_i}_{\bar{\text{B}}},\\
&\ket{1}^{\omega_i}_\text{H}=\ket{1}^{\omega_i}_\text{B}\ket{0}^{\omega_i}_{\bar{\text{B}}},
\end{aligned}
\end{align}
where
\begin{equation}\label{tanq}
\begin{aligned}
\tan\sigma_{d,i}&=\exp{\left(-\frac{\pi \omega_i}{a_0}\right)}=\exp\left(-\frac{\pi\omega_i\sqrt{F_0}}{\sqrt{(1+\ell)}\kappa}\right).
\end{aligned}
\end{equation}

\section{Quantum Entanglement in the Background of a Black Hole: Lorentz Violation Corrections}\label{sec4}
To discuss quantum correlation, assumes Alice and Rob share a maximally entangled Bell state \cite{Pan:2008yr,Wu:2022xwy}, which is expressed as
\begin{align}\label{ball}
\ket{\psi}_\text{AR}=\frac{1}{\sqrt{2}}\left(\ket{0}_\text{AM}\ket{0}_\text{RM}+\ket{1}_\text{AM}\ket{1}_\text{RM}\right),
\end{align}
where Alice's detector is sensitive only to mode $ \ket{n}_\text{AM} $, and Rob's detector is tuned exclusively to mode $ \ket{n}_\text{RM} $.
Given the analogy established earlier between flat spacetime and the black hole scenario, it follows that $ \ket{0}_\text{M} \leftrightarrow \ket{0}_\text{H} $.
Utilizing this correspondence, the maximally entangled state for observers, as they approach the event horizon of a black hole, is represented by
\begin{align}\label{simplemax}
\ket{\psi}_\text{AR}=\frac{1}{\sqrt{2}} (\ket{0}_\text{AH} \ket{0}_\text{RH} + \ket{1}_\text{AH} \ket{1}_\text{RH}).
\end{align}
Here, "A" represents Alice, who is in free fall towards the event horizon, while "R" represents Rob, who remains stationary at a distance $ r=r_0 $ from the black hole.  
Both $ \ket{n}_\text{AH} $ and $ \ket{n}_\text{RH} $ states correspond to number states within the Hartle-Hawking basis.

Following the notation in Eq. \eqref{ABBt}, to analyze the correlations among the bipartite subsystems, we analogously trace out the third subsystem, 
\begin{align}
\begin{aligned}
\rho_\text{AR}=\text{Tr}_{\bar{\text{R}}}\rho_{\text{AR}\bar{\text{R}}},&&
\rho_{\text{A}\bar{\text{R}}}=\text{Tr}_{{\text{R}}}\rho_{\text{AR}\bar{\text{R}}},&&
\rho_{\text{R}\bar{\text{R}}}=\text{Tr}_{{\text{A}}}\rho_{\text{AR}\bar{\text{R}}}.
\end{aligned}
\end{align}
Here, $ \rho_{\text{AR}\bar{\text{R}}} $ is a density operator characterizing a tripartite system, and its representations for the scalar case and the Dirac case are,
\begin{align}
\rho^\text{Scalar}_{\text{AR}\bar{\text{R}}}=&\sum^\infty_{m=0}\langle m\ket{\psi_\text{s}}\langle \psi_\text{s}\ket{m}\non
=&\frac{1}{2\cosh^2\sigma_{s,i}}\sum^\infty_{n=0}\tanh^{2n}\sigma_{s,i}
\Bigg[\ket{0~n~n}\bra{0~n~n} \non
&+\frac{\sqrt{n+1}}{\cosh\sigma_{s,i}}\left( \ket{0~n~n}\bra{1~n+1~n}+\ket{1~n+1~n}\bra{0~n~n} \right)\non
&+\frac{(n+1)}{\cosh^2\sigma_{s,i}} \ket{1~n+1~n}\bra{1~n+1~n} \Bigg],
\end{align}
and
\begin{align}
\rho^\text{Dirac}_{\text{AR}\bar{\text{R}}}=&\ket{\psi_\text{d}}\bra{\psi_\text{d}}\non
=&\frac{1}{2}\Big[\sin\sigma_{d,i}\cos\sigma_{d,i}\left(\ket{0~0~0}\bra{0~1~1}+\ket{0~1~1}\bra{0~0~0} \right)\non
&+\cos\sigma_{d,i}\left(\ket{0~0~0}\bra{1~1~0}+\ket{1~1~0}\bra{0~0~0}  \right)\non
&+\sin\sigma_{d,i}\left(\ket{0~1~1}\bra{1~1~0}+\ket{1~1~0}\bra{0~1~1}  \right)\non
&+\cos^2\sigma_{d,i}\ket{0~0~0}\bra{0~0~0}+\sin^2\sigma_{d,i}\ket{0~1~1}\bra{0~1~1}\non
&+\ket{1~1~0}\bra{1~1~0}\Big],
\end{align}
respectively.
In the AR bipartition, an inertial observer is paired with the field modes available to an accelerated observer. 
The second bipartition, $\text{A}\bar{\text{R}}$, involves Alice and the field modes that Rob is unable to access due to the horizon caused by the gravitational pull of the black hole.
Classical communication is feasible only within the AR and $A\bar{\text{R}}$ bipartitions; these are the only bipartitions where quantum information tasks are feasible \cite{Martin-Martinez:2010yva}.

\subsection{Logarithmic negativity}

In this subsection, we will calculate the logarithmic negativity $ \mathcal{N}(\rho)\equiv \log_2||\rho^\text{T}||_1 $ for the maximally entangled bipartite state \cite{Vidal:2002zz,Diaz:2023jrf,Xu:2024eqg}, for both scalar and Dirac fields.
It is an entanglement monotone sensitive to distillable entanglement, which is defined as the sum of the negative eigenvalues of the partial transpose of the bipartite density matrix, where the partial transpose is taken over only one of the subsystem.
Our primary goal is to compare, side by side, Schwarzschild black holes with those corrected for Lorentz violation, to explore the quantum properties of black holes.

For scalar fields, our focus is on $ \rho_{\text{AR}} $, the density operator characterizing the bipartite system of Alice and Rob.
Alice is located in the asymptotically flat region at infinity, while Rob hovers near the Einstein-Bumblebee black hole. 
They share a physically accessible entanglement, which degrades as Rob's position gets closer to the black hole's event horizon.
We will analyze the entanglement degradation as a function of Rob's position, as described by
\begin{equation}
\begin{aligned}
\mathcal{N}\left(\rho^\text{Scalar}_\text{AR}\right)=&\log_2||\rho^\text{Scalar,T}_\text{AR}||_1\\
=&\log_2\left[\frac{1}{2\cosh^2\sigma_{s,i}}+\sum^\infty_{n=0}\frac{\tanh^{2n}\sigma_{s,i} \sqrt{\mathcal{C}_n}}{2\cosh^2\sigma_{s,i}}  \right],
\end{aligned}
\end{equation}
with
\begin{equation*}
\mathcal{C}_n=\left(\tanh^2\sigma_{s,i}+\frac{n}{\sinh^2\sigma_{s,i}}\right)^2+\frac{4}{\cosh^2\sigma_{s,i}}.
\end{equation*}
If an observer (Rob) is at an infinite distance, $ a(r_0\rightarrow\infty)=0 $, which leads to $ \mathcal{N}(\rho_\text{AR})=1 $. 
When the observer is on the event horizon of the black hole, $ a(r_h)\rightarrow\infty $, and this is identical to the condition $ \sigma_{s,i} \rightarrow \infty $.
Before presenting the results, it is necessary to clearly define the physical quantities under numerical evaluation, which should all be dimensionless. 
Considering the mode frequency measured by Rob and his position measured in black hole radii $R_s$, we define the dimensionless quantities as
\begin{align}
\tilde{\omega}=\omega_i M, && R_0=r_0/R_s.
\end{align}
Eqs. \meq{tansi} and \meq{tanq} can be written as
\begin{align}
\tanh\sigma_{s,i}&=\exp{\left[ -4\pi \tilde{\omega}\sqrt{1+\ell-\frac{1+\ell}{R_0}} \right]},
\end{align}
\begin{align}
\tan\sigma_{d,i}&=\exp{\left[ -4\pi \tilde{\omega}\sqrt{1+\ell-\frac{1+\ell}{R_0}} \right]}.
\end{align}
\begin{figure}[h]
\centering 
\includegraphics[width=0.48\linewidth]{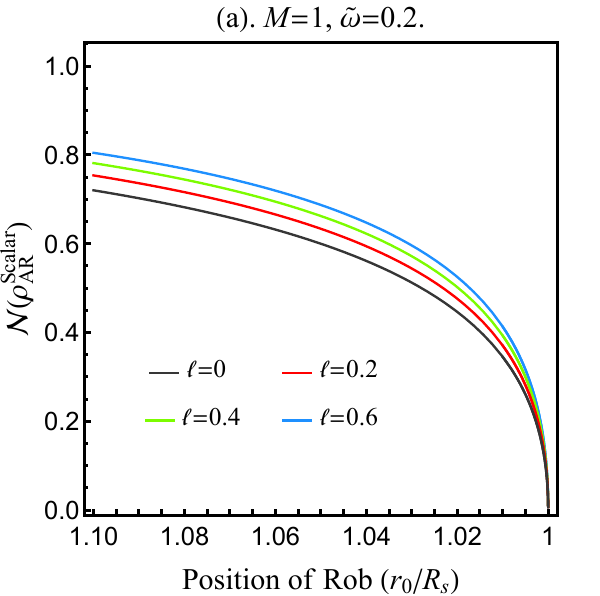} 
\includegraphics[width=0.48\linewidth]{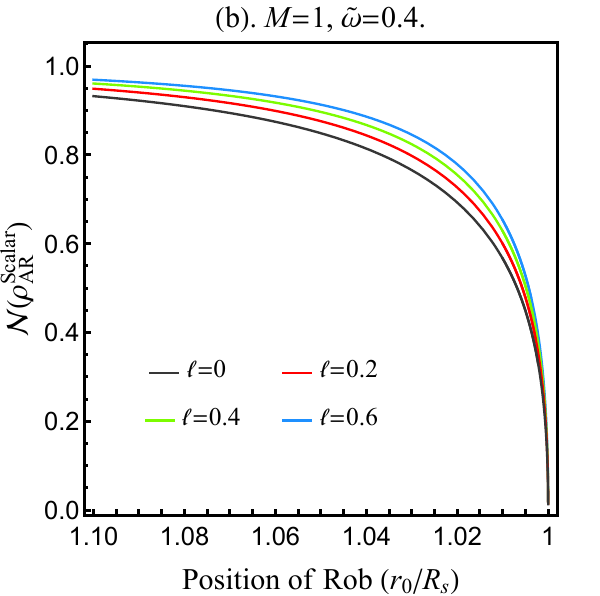}  
\includegraphics[width=0.48\linewidth]{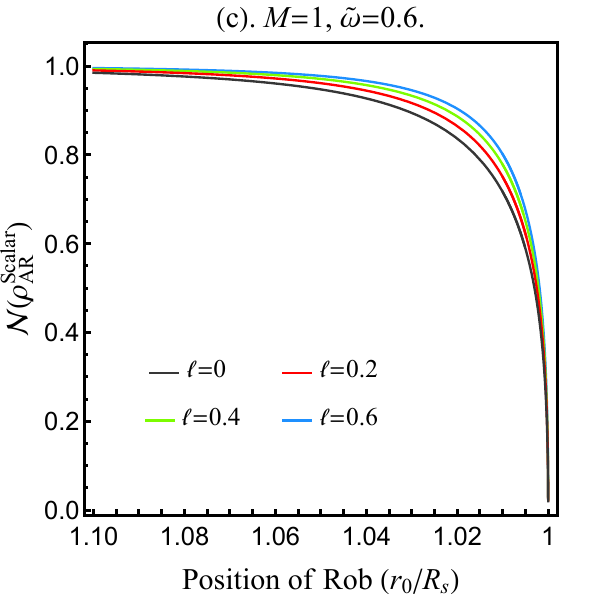}  
\includegraphics[width=0.48\linewidth]{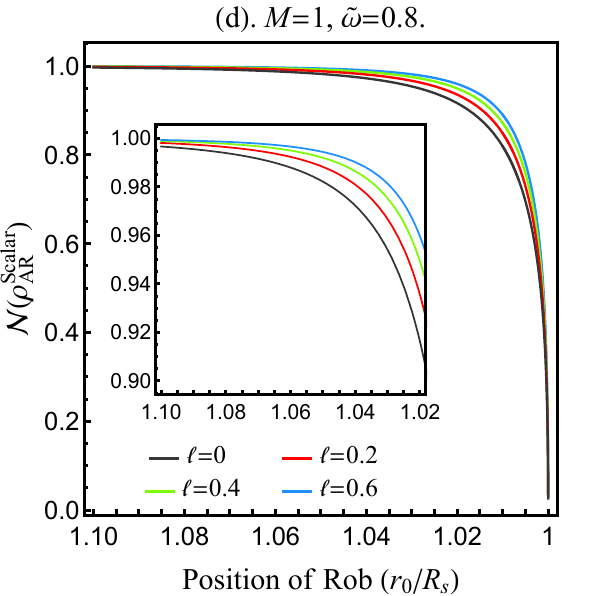}   
\caption{
Scalar field: The entanglement of the Alice-Rob system is considered as a function of Rob's position for different values of $ \ell $. The entanglement vanishes as Rob approaches the black hole radius, and no entanglement is created between Alice and antiRob. 
The smaller the value of $ \ell $ or $ \tilde{\omega} $, the more degradation is caused by the black hole.}
\label{fig1}
\end{figure}
Here, to visualize the deviations introduced by the corrections, we selected relatively extreme values for the Lorentz violation parameters.
Figs. \ref{fig1} show that the phenomenon of entanglement degradation is universal across different mode frequencies, but it is also influenced by Lorentz violation, causing deviations from the behavior observed in Schwarzschild black holes.
In other words, the coupling between spacetime and the Lorentz-violating vector field leads to a weakening of gravity-induced entanglement degradation. 
At sufficiently high mode frequencies, this weakening becomes negligible (unless very close to the event horizon).
However, at sufficiently low mode frequencies, the effects of Lorentz violation become highly pronounced. 
This suggests that investigating Lorentz violation at astrophysical scales relies heavily on low-frequency detectors. 
Since the energy of low-frequency fields is relatively low, the non-zero VEV of the bumblebee field will contribute more significantly.
As Rob approaches the event horizon, the entanglement between Alice and Rob diminishes, eventually vanishing entirely when Rob gets extremely close to the horizon. 
The entanglement degradation takes place in a narrow region near the event horizon, and the presence of a Lorentz-violating parameter further constricts this region.


\begin{figure}[h]
\centering 
\includegraphics[width=0.48\linewidth]{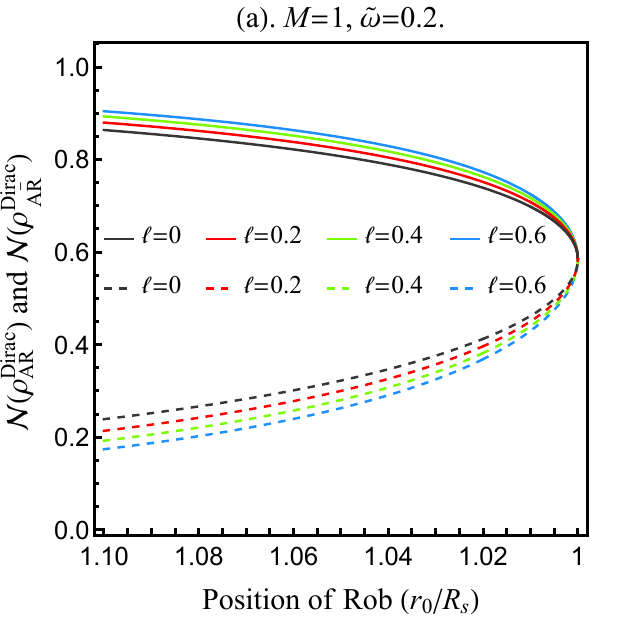} 
\includegraphics[width=0.48\linewidth]{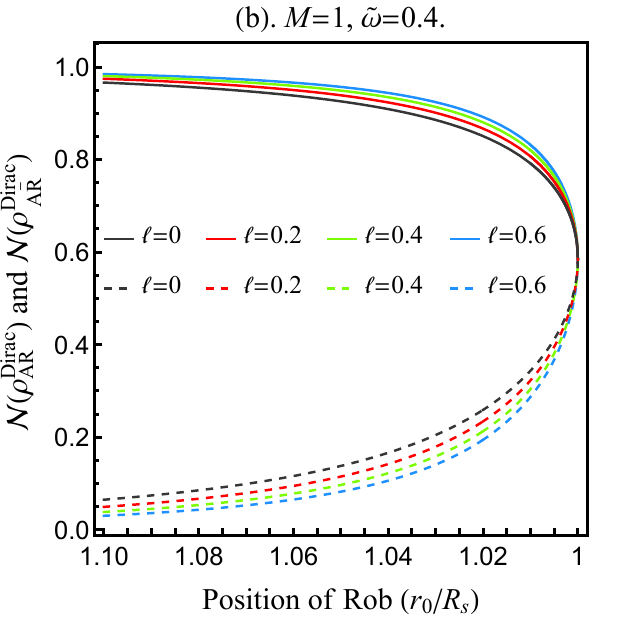}  
\includegraphics[width=0.48\linewidth]{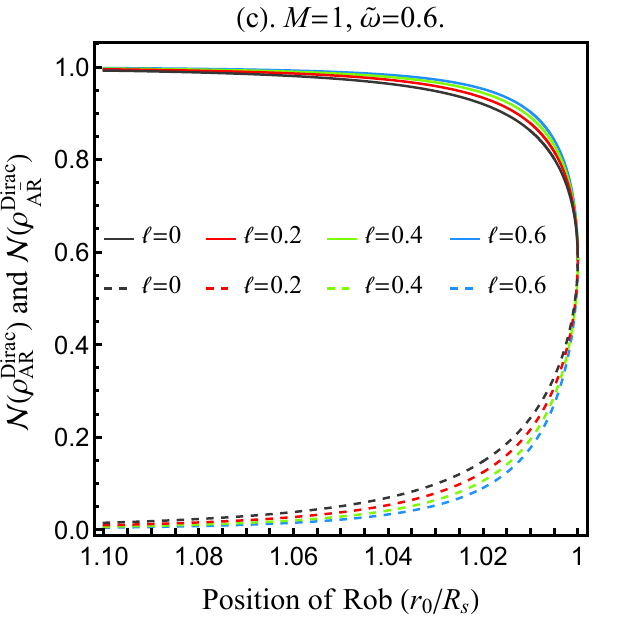}  
\includegraphics[width=0.48\linewidth]{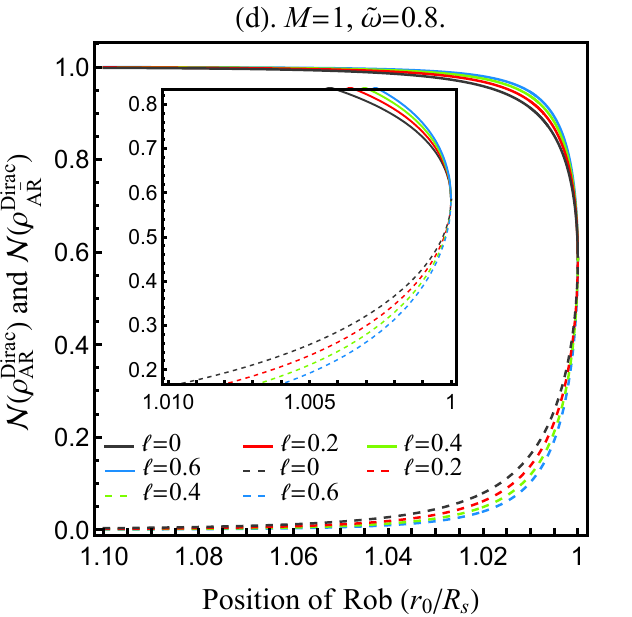}   
\caption{Dirac field: The entanglement of the Alice-Rob system (solid line) and the Alice-antiRob system (dashed line) as a function of Rob's position for different values of $ \ell $.
When $ \ell $ is large, the degradation of entanglement in AR slows down.
The degradation does not reach its maximum, and its value remains independent of $ \ell $ and $ \tilde{\omega} $.}
\label{fig2}
\end{figure}
For Dirac fields, due to the physical characteristics of entanglement redistribution \cite{Wang:2010bf}, we focus on both $ \rho_{\text{AR}} $ and $ \rho_{\text{A}\bar{\text{R}}} $, the density operators characterizing the bipartite systems of Alice-Rob and Alice-antiRob, respectively.
The logarithmic negativity in this case are give by:
\begin{equation}
\begin{aligned}
\mathcal{N}\left(\rho^\text{Dirac}_\text{AR}\right)=&\log_2||\rho^\text{Dirac,T}_\text{AR}||_1=\log_2\left(1+\cos^2\sigma_{d,i}\right),
\end{aligned}
\end{equation}
\begin{equation}
\begin{aligned}
\mathcal{N}\left(\rho^\text{Dirac}_{\text{A}\bar{\text{R}}}\right)=&\log_2||\rho^\text{Dirac,T}_{\text{A}\bar{\text{R}}}||_1=\log_2\left(1+\sin^2\sigma_{d,i}\right).
\end{aligned}
\end{equation}
Figs. \ref{fig2} show the Dirac fields, something very different happens.
We observe that, in the bipartition AR, the correlations decrease to a specific finite limit $ \mathcal{N}\simeq0.58 $ for any given Lorentz violation parameter $\ell$ and mode frequencies $\tilde{\omega}$. 
This indicates that entanglement persists even as Rob approaches the event horizon asymptotically, a well-known phenomenon in both Rindler and Schwarzschild scenarios \cite{Alsing:2006cj,Ahn:2008zf}.
\begin{figure}[h]
\centering 
\includegraphics[width=0.7\linewidth]{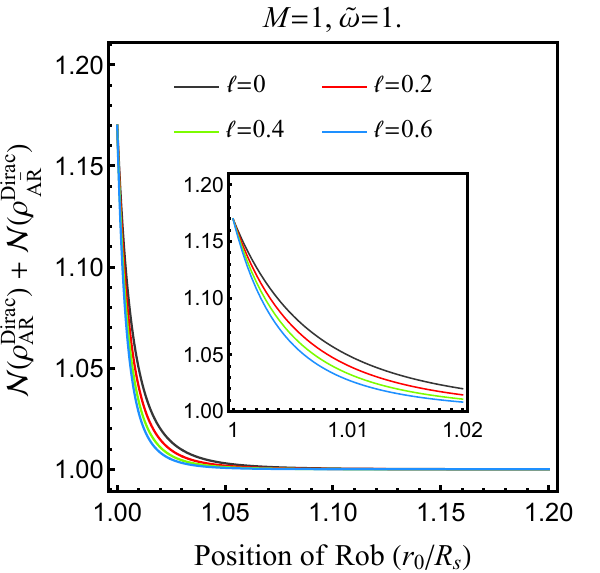}   
\caption{ Driac field: The entanglement sum of the Alice-Rob system and the Alice-antiRob system, $ \mathcal{N}(\rho_{\text{A}{\text{R}}})+\mathcal{N}(\rho_{\text{A}\bar{\text{R}}}) $, is shown as a function of Rob's position for different values of $ \ell $. }
\label{fig2.1}
\end{figure}
At the same time that entanglement is destroyed in the AR bipartition, it is created in the complementary $A\bar{\text{R}}$ bipartition.
When Rob crosses the event horizon, Rob and antiRob coincide, and the bipartition $A\bar{\text{R}}$ exhibits maximum entanglement with $ \mathcal{N}(\rho_{\text{A}\bar{\text{R}}})\simeq0.58 $. 
This means that the logarithmic negativity measure of entanglement does not follow a conservation law at certain positions outside the horizon, when $ r_0/R_s\rightarrow 1 $ lead $ \mathcal{N}(\rho_{\text{A}{\text{R}}})+\mathcal{N}(\rho_{\text{A}\bar{\text{R}}})>1 $, as shown in Fig. \ref{fig2.1}.


Based on Figs.\ref{fig1} and \ref{fig2}, we can also conclude that all relevant entanglement degradation phenomena occur near the event horizon of the bumblebee black hole, even under more extreme conditions where we apply Lorentz-violating modifications to the spacetime. 
Therefore, the Rindler approximation we consider is valid, and the event horizon is not expected to affect the entangled system at distances far from it.

\subsection{Mutual information}\label{sec5}

In order to discuss this further, we will calculate the mutual information for the maximally entangled bipartite state, for both scalar and Dirac fields.
The mutual information, which gives an idea of the total amount of correlation, is defined for a bipartite system AR as \cite{Fuentes-Schuller:2004iaz}
\begin{align}
I(\rho_\text{AR})=S(\rho_\text{A})+S(\rho_\text{R})-S(\rho_\text{AR}),
\end{align}
where $ S(\rho)=-\text{Tr}(\rho \log_2\rho) =-\sum_i\lambda_{i}\log_2\lambda_{i}$ is the entropy of the density matrix $ \rho $, with $\lambda_{i}$ being its eigenvalue.

For scalar fields, the entropy of the joint state is
\begin{equation}
\begin{aligned}
S(\rho^\text{Scalar}_\text{AR})=&-\sum_{n=0}^{\infty}\frac{\tanh^{2n}\sigma_{s,i}}{2\cosh^2\sigma_{s,i}}\left(1+\frac{n+1}{\cosh^2\sigma_{s,i}}\right)\\
&\times\log_2\left[\frac{\tanh^{2n}\sigma_{s,i}}{2\cosh^2\sigma_{s,i}}\left(1+\frac{n+1}{\cosh^2\sigma_{s,i}}\right)\right].
\end{aligned}
\end{equation}
The density matrix for Rob is obtained by tracing out Alice's states, and its entropy is
\begin{equation}
\begin{aligned}
S(\rho^\text{Scalar}_\text{R})=&-\sum_{n=0}^{\infty}\frac{\tanh^{2n}\sigma_{s,i}}{2\cosh^2\sigma_{s,i}}\left(1+\frac{n}{\sinh^2\sigma_{s,i}}\right)\\
&\times\log_2\left[\frac{\tanh^{2n}\sigma_{s,i}}{2\cosh^2\sigma_{s,i}}\left(1+\frac{n}{\sinh^2\sigma_{s,i}}\right)\right].
\end{aligned}
\end{equation}
Given that $ S(\rho^\text{Scalar}_\text{A})=1 $, the mutual information is
\begin{equation}
\begin{aligned}
I(\rho_\text{AR}^\text{Scalar})=&1-\frac{1}{2}\log_2\tanh^2\sigma_{s,i}-\sum_{n=0}^{\infty}\frac{\tanh^{2n}\sigma_{s,i}}{2\cosh^2\sigma_{s,i}} \bar{\mathcal{C}}_n
\end{aligned}
\end{equation}
with
\begin{align*}
\bar{\mathcal{C}}_n=&\left(1+\frac{n}{\sinh^2\sigma_{s,i}}\right)\log_2\left(1+\frac{n}{\sinh^2\sigma_{s,i}}\right)\\
&-\left(1+\frac{n+1}{\cosh^2\sigma_{s,i}}\right)\log_2\left(1+\frac{n+1}{\cosh^2\sigma_{s,i}}\right),
\end{align*}
which we plot in Figs. \ref{fig3}.
\textbf{\begin{figure}[h]
\centering 
\includegraphics[width=0.48\linewidth]{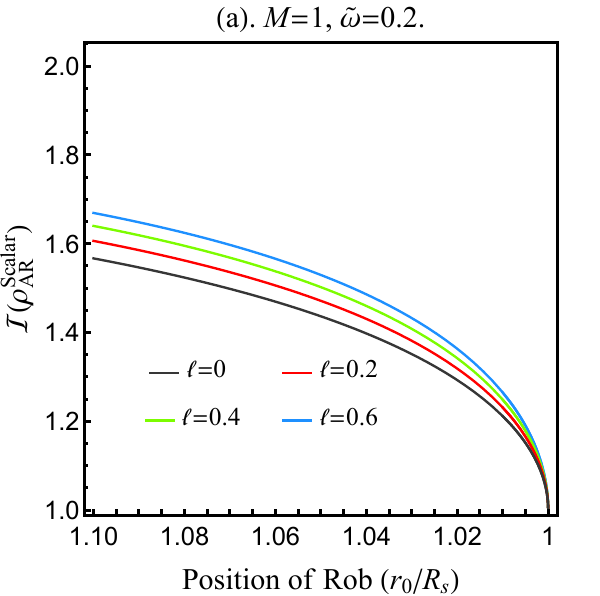} 
\includegraphics[width=0.48\linewidth]{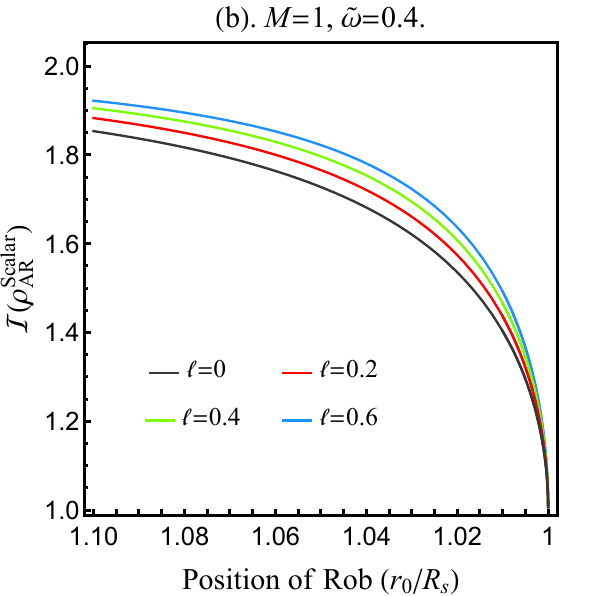}  
\includegraphics[width=0.48\linewidth]{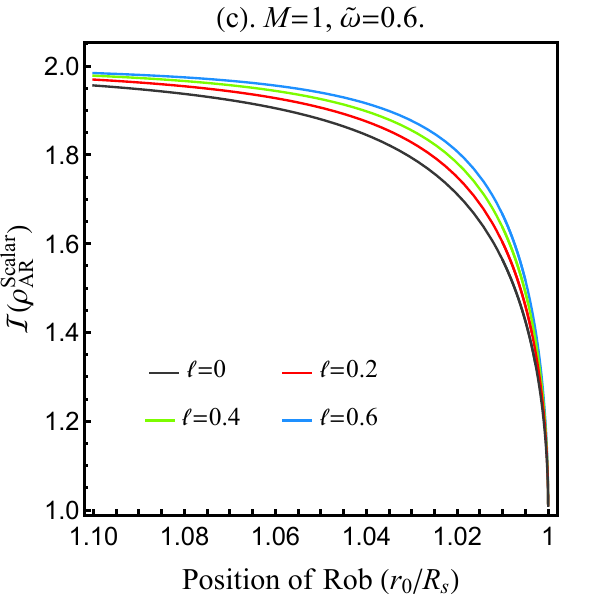}  
\includegraphics[width=0.48\linewidth]{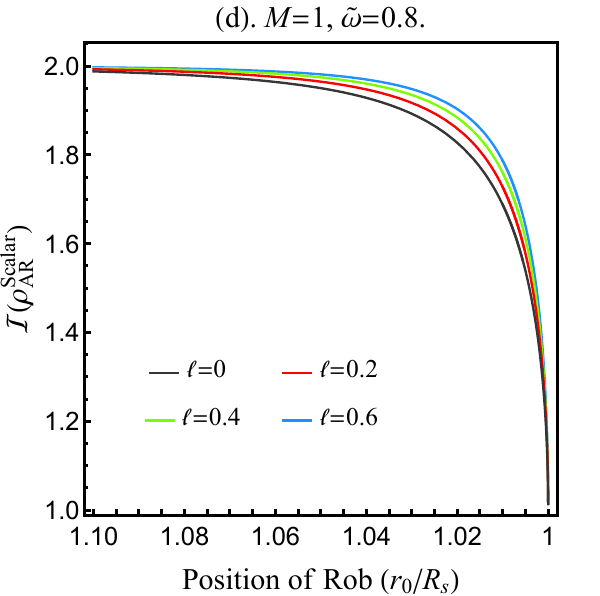}   
\caption{ Scalar field: The mutual information of the Alice-Rob system is analyzed as a function of Rob's position for different values of $ \ell $.}
\label{fig3}
\end{figure}}
When Rob's position is far from the event horizon of the bumblebee black hole, the mutual information is 2. 
As Rob approaches the event horizon, it becomes smaller and converges to unity as Rob enters the event horizon. 
It is important to notice that for the Lorentz-violating corrections, the mutual information degrades at a slower rate, which is consistent with the conclusion in the previous section. 
When the mutual information becomes unity, there is no distillable entanglement left between the two states.

For Dirac fields, the mutual information of the bipartite systems AR and $ \text{A}\bar{\text{R}} $ are given by
\begin{equation}
\begin{aligned}
I(\rho^\text{Dirac}_{\text{A}\text{R}})=&1-\frac{1}{2}\cos^2\sigma_{d,i}\log_2\left(\frac{\cos^2\sigma_{d,i}}{2}\right)\\
&-\left(1-\frac{1}{2}\cos^2\sigma_{d,i}\right)\log_2\left(1-\frac{1}{2}\cos^2\sigma_{d,i}\right)\\
&+\frac{1}{2}\left(1+\cos^2\sigma_{d,i}\right)\log_2\left(\frac{1+\cos^2\sigma_{d,i}}{2}\right)\\
&+\frac{1}{2}\left(1-\cos^2\sigma_{d,i}\right)\log_2\left( \frac{1-\cos^2\sigma_{d,i}}{2} \right),
\end{aligned}
\end{equation}
and
\begin{equation}
\begin{aligned}
I(\rho^\text{Dirac}_{\text{A}\bar{\text{R}}})=&1-\frac{1}{2}\sin^2\sigma_{d,i}\log_2\left(\frac{\sin^2\sigma_{d,i}}{2}\right)\\
&-\left(1-\frac{1}{2}\sin^2\sigma_{d,i}\right)\log_2\left(1-\frac{1}{2}\sin^2\sigma_{d,i}\right)\\
&+\frac{1}{2}\left(1+\sin^2\sigma_{d,i}\right)\log_2\left(\frac{1+\sin^2\sigma_{d,i}}{2}\right)\\
&+\frac{1}{2}\left(1-\sin^2\sigma_{d,i}\right)\log_2\left( \frac{1-\sin^2\sigma_{d,i}}{2} \right),
\end{aligned}
\end{equation}
respectively.

\begin{figure}[h]
\centering 
\includegraphics[width=0.48\linewidth]{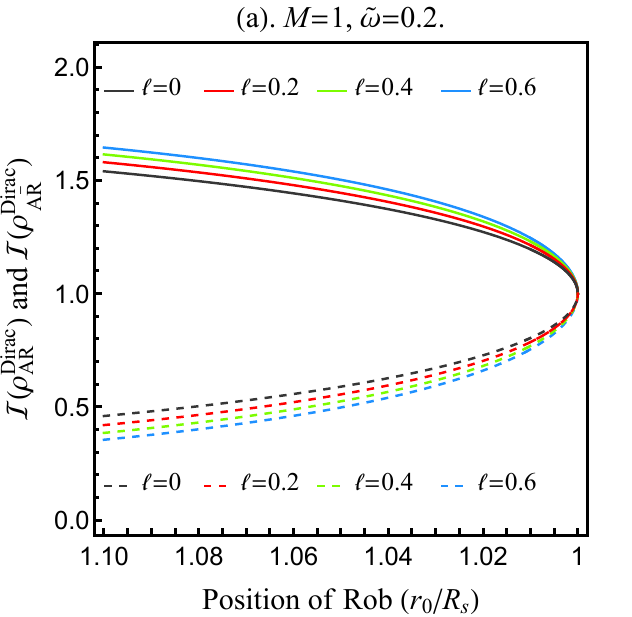} 
\includegraphics[width=0.48\linewidth]{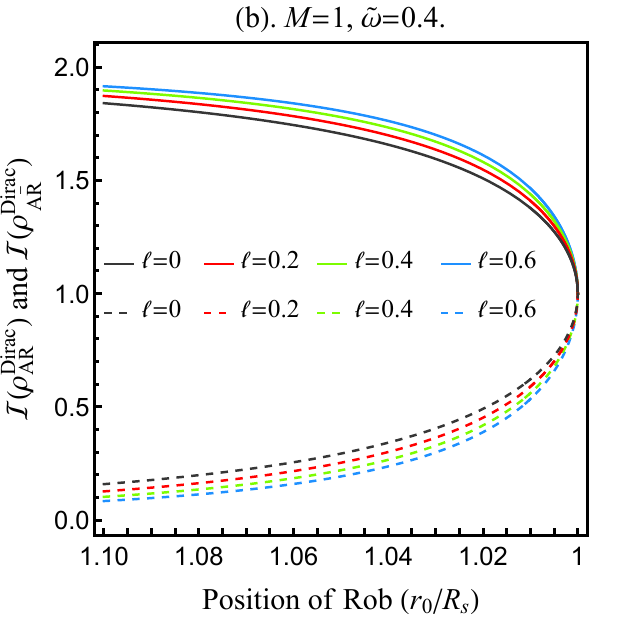}  
\includegraphics[width=0.48\linewidth]{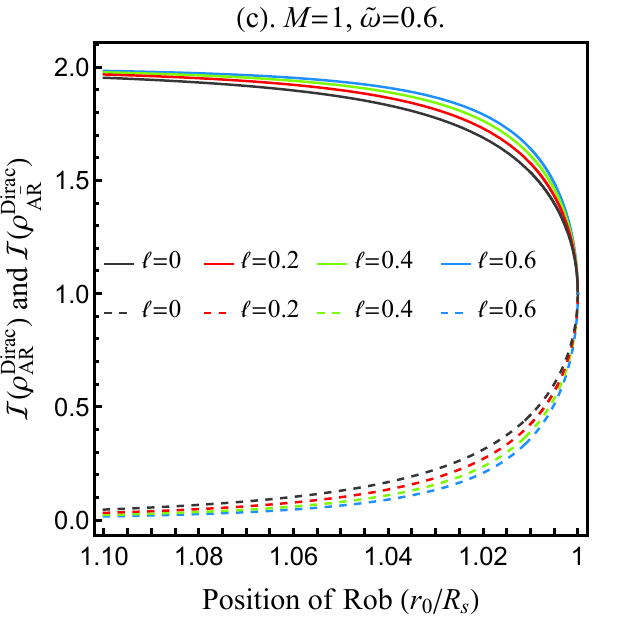}  
\includegraphics[width=0.48\linewidth]{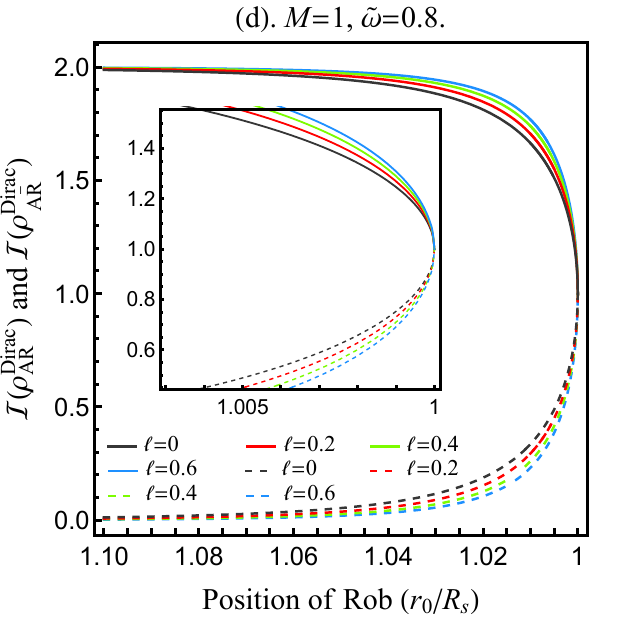}   
\caption{
Dirac field: The mutual information of the Alice-Rob system (solid line) and the Alice-antiRob system (dashed line) as a function of Rob's position for different values of $ \ell $. 
Mutual information AR decreases as Rob is closer to the horizon and mutual information $ \text{A}\bar{\text{R}} $ grows.
When $ \ell $ is large, the degradation of mutual information in the Alice-Rob (antiRob) system slows down (up).}
\label{fig4}
\end{figure}

The results for the Dirac fields are shown in Figs. \ref{fig4}.
We can observe that the behavior of the mutual information is very similar to the logarithmic negativity discussed in the previous section, and the effects of the Lorentz violation are also consistent. 
However, unlike logarithmic negativity, the mutual information conservation law generally holds, which was first discovered in Rindler spacetime \cite{Martin-Martinez:2010bcj}.
Specifically, in the bumblebee black hole, this conservation law is maintained and is not affected by Lorentz violation.
Namely, for any distance to the horizon or Lorenz violates the parameter it is fulfilled that
\begin{align}
I_{\text{AR}}+I_{{\text{A}\bar{\text{R}}}}=2.
\end{align}

\section{Conclusions}

In this manuscript, we investigate the phenomenon of entanglement degradation for a black hole coupled to a Lorentz-violating vector field, ie,eg., Einstein-Bumblebee black hole.
In the near-horizon approximation, the Einstein-Bumblebee black hole metric can be expressed in the Rindler form.
This formulation not only facilitates the identification of three timelike Killing vectors but also aids in recognizing the vacuum modes and their analogies to those in flat spacetime, a tool originally developed by E. Martin-Martinez and colleagues in their study of entanglement degradation for uniformly accelerated observers.
The Rindler limit of infinite acceleration can reproduces the scenario where the black hole is coupled to a Lorentz-violating vector field, in which Rob is arbitrarily close to the event horizon.
By carefully analyzing the fine structure of this limit, we can explicitly reveal how the entanglement degradation phenomenon depends on the distance to the horizon, the Einstein-Bumblebee black hole parameters, and the mode frequency $ \omega $ of the entangled mode under consideration, while ensuring that the approximation remains accurate enough for the toolbox developed for the Rindler case to be rigorously applied here.

We considered both scalar and Dirac fields, using logarithmic negativity and mutual information as measures of entanglement, respectively.
Unlike the Schwarzschild case, the Lorentz violation parameter $ \ell $ affects entanglement with the black hole, leading to important results.
We observed the following trends:
\begin{itemize}[leftmargin=*,noitemsep,topsep=0pt,partopsep=0pt]
\item 
\textbf{Scalar field:}
As Rob approaches the black hole event horizon, the entanglement of the AR bipartite system will degrade, and this degradation process will slow down due to the non-minimally coupled Lorentz-violating vector field of gravitational origin.
The entanglement degradation occurs in a narrow region near the event horizon, and a Lorentz-violating parameter further constricts this region.
Moreover, the logarithmic negativity and mutual information exhibit the same behavior in relation to entanglement degradation under different Lorentz-violating parameters.
\item 
\textbf{Driac field:} 
The Lorentz violation parameter exhibits an effect on the bipartite system AR similar to that of a scalar field. 
The difference is that the entanglement does not completely degrade until Rob falls into the black hole's event horizon.
For this field, the mutual information satisfies the conservation law, whether or not Lorentz violation exists in spacetime, near the event horizon.
\end{itemize}

The analysis method in this paper is also applicable to models where a black hole is coupled to a Lorentz-violating tensor field \cite{Kalb:1974yc,Yang:2023wtu,Duan:2023gng,Liu:2024oas,Liu:2024lve}. 
In such cases, the spacetime structure will differ from the scenario described by equation \meq{ds2MOG}, as Lorentz violation affects the location of the black hole's event horizon, causing it to deviate from the Schwarzschild radius. 
By comparing the phenomena of entanglement degradation in different fields within Lorentz-violating spacetimes, induced by the coupling of vectors and tensors, we can gain a deeper insight into the quantum properties of spacetime.
Moreover, Hawking radiation not only leads to entanglement degradation but also carries entangled states generated near the black hole horizon during the evaporation process. 
The Unruh-DeWitt detector model can simulate the interaction between quantum detectors and quantum fields, showing that even detectors near the black hole horizon can harvest entanglement from quantum fields outside the black hole \cite{Henderson:2017yuv,Cong:2020nec,Liu:2022uhf,Maeso-Garcia:2022uzf,Liu:2023awu,Liu:2023zro,Lindel:2023rfi,Ji:2024fcq,Wu:2024whx,Zeng:2024qme,Liu:2024pse}, such as the quantum field associated with Hawking radiation.
Future research on the impact of Lorentz violations on entanglement capture could provide valuable perspectives on quantum information in black hole spacetimes, revealing new aspects of how spacetime and quantum phenomena interact under these conditions.



\acknowledgments
We are greatly indebted to the anonymous referee for constructive comments, which improved the presentation of this work. 
The authors also gratefully acknowledge Shu-Min Wu for insightful discussions.
This work was supported by the National Natural Science Foundation of China under Grants  No. 12475051; the innovative research group of Hunan Province under Grant No. 2024JJ1006; the Natural Science Foundation of Hunan Province under grant No. 2023JJ30384; and the science and technology innovation Program of Hunan Province under grant No. 2024RC1050.

\appendix
%


\begin{thebibliography}{78}%
\makeatletter
\providecommand \@ifxundefined [1]{%
 \@ifx{#1\undefined}
}%
\providecommand \@ifnum [1]{%
 \ifnum #1\expandafter \@firstoftwo
 \else \expandafter \@secondoftwo
 \fi
}%
\providecommand \@ifx [1]{%
 \ifx #1\expandafter \@firstoftwo
 \else \expandafter \@secondoftwo
 \fi
}%
\providecommand \natexlab [1]{#1}%
\providecommand \enquote  [1]{``#1''}%
\providecommand \bibnamefont  [1]{#1}%
\providecommand \bibfnamefont [1]{#1}%
\providecommand \citenamefont [1]{#1}%
\providecommand \href@noop [0]{\@secondoftwo}%
\providecommand \href [0]{\begingroup \@sanitize@url \@href}%
\providecommand \@href[1]{\@@startlink{#1}\@@href}%
\providecommand \@@href[1]{\endgroup#1\@@endlink}%
\providecommand \@sanitize@url [0]{\catcode `\\12\catcode `\$12\catcode
  `\&12\catcode `\#12\catcode `\^12\catcode `\_12\catcode `\%12\relax}%
\providecommand \@@startlink[1]{}%
\providecommand \@@endlink[0]{}%
\providecommand \url  [0]{\begingroup\@sanitize@url \@url }%
\providecommand \@url [1]{\endgroup\@href {#1}{\urlprefix }}%
\providecommand \urlprefix  [0]{URL }%
\providecommand \Eprint [0]{\href }%
\providecommand \doibase [0]{https://doi.org/}%
\providecommand \selectlanguage [0]{\@gobble}%
\providecommand \bibinfo  [0]{\@secondoftwo}%
\providecommand \bibfield  [0]{\@secondoftwo}%
\providecommand \translation [1]{[#1]}%
\providecommand \BibitemOpen [0]{}%
\providecommand \bibitemStop [0]{}%
\providecommand \bibitemNoStop [0]{.\EOS\space}%
\providecommand \EOS [0]{\spacefactor3000\relax}%
\providecommand \BibitemShut  [1]{\csname bibitem#1\endcsname}%
\let\auto@bib@innerbib\@empty
\bibitem [{\citenamefont {Boschi}\ \emph {et~al.}(1998)\citenamefont {Boschi},
  \citenamefont {Branca}, \citenamefont {De~Martini}, \citenamefont {Hardy},\
  and\ \citenamefont {Popescu}}]{Boschi:1997dg}%
  \BibitemOpen
  \bibfield  {author} {\bibinfo {author} {\bibfnamefont {D.}~\bibnamefont
  {Boschi}}, \bibinfo {author} {\bibfnamefont {S.}~\bibnamefont {Branca}},
  \bibinfo {author} {\bibfnamefont {F.}~\bibnamefont {De~Martini}}, \bibinfo
  {author} {\bibfnamefont {L.}~\bibnamefont {Hardy}},\ and\ \bibinfo {author}
  {\bibfnamefont {S.}~\bibnamefont {Popescu}},\ }\bibfield  {title} {\bibinfo
  {title} {{Experimental realization of teleporting an unknown pure quantum
  state via dual classical and Einstein-Podolski-Rosen channels}},\ }\href
  {https://doi.org/10.1103/PhysRevLett.80.1121} {\bibfield  {journal} {\bibinfo
   {journal} {Phys. Rev. Lett.}\ }\textbf {\bibinfo {volume} {80}},\ \bibinfo
  {pages} {1121} (\bibinfo {year} {1998})},\ \Eprint
  {https://arxiv.org/abs/9710013} {arXiv:9710013 [quant-ph]} \BibitemShut
  {NoStop}%
\bibitem [{\citenamefont {Pan}\ \emph {et~al.}(2001)\citenamefont {Pan},
  \citenamefont {Simon}, \citenamefont {Brukner},\ and\ \citenamefont
  {Zeilinger}}]{Pan:2001nrz}%
  \BibitemOpen
  \bibfield  {author} {\bibinfo {author} {\bibfnamefont {J.~W.}\ \bibnamefont
  {Pan}}, \bibinfo {author} {\bibfnamefont {C.}~\bibnamefont {Simon}}, \bibinfo
  {author} {\bibfnamefont {v.}~\bibnamefont {Brukner}},\ and\ \bibinfo {author}
  {\bibfnamefont {A.}~\bibnamefont {Zeilinger}},\ }\bibfield  {title} {\bibinfo
  {title} {{Entanglement purification for quantum communication}},\ }\href
  {https://doi.org/10.1038/35074041} {\bibfield  {journal} {\bibinfo  {journal}
  {Nature}\ }\textbf {\bibinfo {volume} {410}},\ \bibinfo {pages} {1067}
  (\bibinfo {year} {2001})}\BibitemShut {NoStop}%
\bibitem [{\citenamefont {Hu}\ \emph {et~al.}(2018)\citenamefont {Hu},
  \citenamefont {Hu}, \citenamefont {Wang}, \citenamefont {Peng}, \citenamefont
  {Zhang},\ and\ \citenamefont {Fan}}]{Hu:2018mni}%
  \BibitemOpen
  \bibfield  {author} {\bibinfo {author} {\bibfnamefont {M.-L.}\ \bibnamefont
  {Hu}}, \bibinfo {author} {\bibfnamefont {X.}~\bibnamefont {Hu}}, \bibinfo
  {author} {\bibfnamefont {J.}~\bibnamefont {Wang}}, \bibinfo {author}
  {\bibfnamefont {Y.}~\bibnamefont {Peng}}, \bibinfo {author} {\bibfnamefont
  {Y.-R.}\ \bibnamefont {Zhang}},\ and\ \bibinfo {author} {\bibfnamefont
  {H.}~\bibnamefont {Fan}},\ }\bibfield  {title} {\bibinfo {title} {{Quantum
  coherence and geometric quantum discord}},\ }\href
  {https://doi.org/10.1016/j.physrep.2018.07.004} {\bibfield  {journal}
  {\bibinfo  {journal} {Phys. Rept.}\ }\textbf {\bibinfo {volume} {762-764}},\
  \bibinfo {pages} {1} (\bibinfo {year} {2018})}\BibitemShut {NoStop}%
\bibitem [{\citenamefont {Fuentes-Schuller}\ and\ \citenamefont
  {Mann}(2005)}]{Fuentes-Schuller:2004iaz}%
  \BibitemOpen
  \bibfield  {author} {\bibinfo {author} {\bibfnamefont {I.}~\bibnamefont
  {Fuentes-Schuller}}\ and\ \bibinfo {author} {\bibfnamefont {R.~B.}\
  \bibnamefont {Mann}},\ }\bibfield  {title} {\bibinfo {title} {{Alice falls
  into a black hole: Entanglement in non-inertial frames}},\ }\href
  {https://doi.org/10.1103/PhysRevLett.95.120404} {\bibfield  {journal}
  {\bibinfo  {journal} {Phys. Rev. Lett.}\ }\textbf {\bibinfo {volume} {95}},\
  \bibinfo {pages} {120404} (\bibinfo {year} {2005})},\ \Eprint
  {https://arxiv.org/abs/0410172} {arXiv:0410172 [quant-ph]} \BibitemShut
  {NoStop}%
\bibitem [{\citenamefont {Alsing}\ \emph {et~al.}(2006)\citenamefont {Alsing},
  \citenamefont {Fuentes-Schuller}, \citenamefont {Mann},\ and\ \citenamefont
  {Tessier}}]{Alsing:2006cj}%
  \BibitemOpen
  \bibfield  {author} {\bibinfo {author} {\bibfnamefont {P.~M.}\ \bibnamefont
  {Alsing}}, \bibinfo {author} {\bibfnamefont {I.}~\bibnamefont
  {Fuentes-Schuller}}, \bibinfo {author} {\bibfnamefont {R.~B.}\ \bibnamefont
  {Mann}},\ and\ \bibinfo {author} {\bibfnamefont {T.~E.}\ \bibnamefont
  {Tessier}},\ }\bibfield  {title} {\bibinfo {title} {{Entanglement of Dirac
  fields in non-inertial frames}},\ }\href
  {https://doi.org/10.1103/PhysRevA.74.032326} {\bibfield  {journal} {\bibinfo
  {journal} {Phys. Rev. A}\ }\textbf {\bibinfo {volume} {74}},\ \bibinfo
  {pages} {032326} (\bibinfo {year} {2006})},\ \Eprint
  {https://arxiv.org/abs/0603269} {arXiv:0603269 [quant-ph]} \BibitemShut
  {NoStop}%
\bibitem [{\citenamefont {Wang}\ \emph
  {et~al.}(2010{\natexlab{a}})\citenamefont {Wang}, \citenamefont {Deng},\ and\
  \citenamefont {Jing}}]{Wang:2009qg}%
  \BibitemOpen
  \bibfield  {author} {\bibinfo {author} {\bibfnamefont {J.}~\bibnamefont
  {Wang}}, \bibinfo {author} {\bibfnamefont {J.}~\bibnamefont {Deng}},\ and\
  \bibinfo {author} {\bibfnamefont {J.}~\bibnamefont {Jing}},\ }\bibfield
  {title} {\bibinfo {title} {{Classical correlation and quantum discord sharing
  of Dirac fields in noninertial frames}},\ }\href
  {https://doi.org/10.1103/PhysRevA.81.052120} {\bibfield  {journal} {\bibinfo
  {journal} {Phys. Rev. A}\ }\textbf {\bibinfo {volume} {81}},\ \bibinfo
  {pages} {052120} (\bibinfo {year} {2010}{\natexlab{a}})},\ \Eprint
  {https://arxiv.org/abs/0912.4129} {arXiv:0912.4129 [quant-ph]} \BibitemShut
  {NoStop}%
\bibitem [{\citenamefont {Martin-Martinez}\ and\ \citenamefont
  {Leon}(2009)}]{Martin-Martinez:2009hfq}%
  \BibitemOpen
  \bibfield  {author} {\bibinfo {author} {\bibfnamefont {E.}~\bibnamefont
  {Martin-Martinez}}\ and\ \bibinfo {author} {\bibfnamefont {J.}~\bibnamefont
  {Leon}},\ }\bibfield  {title} {\bibinfo {title} {{Are Alice and Rob really
  protected by statistics as she falls into a black hole?}},\ }\href
  {https://doi.org/10.1103/PhysRevA.80.042318} {\bibfield  {journal} {\bibinfo
  {journal} {Phys. Rev. A}\ }\textbf {\bibinfo {volume} {80}},\ \bibinfo
  {pages} {042318} (\bibinfo {year} {2009})},\ \Eprint
  {https://arxiv.org/abs/0907.1960} {arXiv:0907.1960 [quant-ph]} \BibitemShut
  {NoStop}%
\bibitem [{\citenamefont {Martin-Martinez}\ \emph {et~al.}(2010)\citenamefont
  {Martin-Martinez}, \citenamefont {Garay},\ and\ \citenamefont
  {Leon}}]{Martin-Martinez:2010yva}%
  \BibitemOpen
  \bibfield  {author} {\bibinfo {author} {\bibfnamefont {E.}~\bibnamefont
  {Martin-Martinez}}, \bibinfo {author} {\bibfnamefont {L.~J.}\ \bibnamefont
  {Garay}},\ and\ \bibinfo {author} {\bibfnamefont {J.}~\bibnamefont {Leon}},\
  }\bibfield  {title} {\bibinfo {title} {{Unveiling quantum entanglement
  degradation near a Schwarzschild black hole}},\ }\href
  {https://doi.org/10.1103/PhysRevD.82.064006} {\bibfield  {journal} {\bibinfo
  {journal} {Phys. Rev. D}\ }\textbf {\bibinfo {volume} {82}},\ \bibinfo
  {pages} {064006} (\bibinfo {year} {2010})},\ \Eprint
  {https://arxiv.org/abs/1006.1394} {arXiv:1006.1394 [quant-ph]} \BibitemShut
  {NoStop}%
\bibitem [{\citenamefont {Esfahani}\ \emph {et~al.}(2011)\citenamefont
  {Esfahani}, \citenamefont {Shamirzai},\ and\ \citenamefont
  {Soltani}}]{Esfahani:2010gt}%
  \BibitemOpen
  \bibfield  {author} {\bibinfo {author} {\bibfnamefont {B.~N.}\ \bibnamefont
  {Esfahani}}, \bibinfo {author} {\bibfnamefont {M.}~\bibnamefont
  {Shamirzai}},\ and\ \bibinfo {author} {\bibfnamefont {M.}~\bibnamefont
  {Soltani}},\ }\bibfield  {title} {\bibinfo {title} {{Reduction of
  entanglement degradation and teleportation improvement in
  Einstein-Gauss-Bonnet gravity}},\ }\href
  {https://doi.org/10.1103/PhysRevD.84.025024} {\bibfield  {journal} {\bibinfo
  {journal} {Phys. Rev. D}\ }\textbf {\bibinfo {volume} {84}},\ \bibinfo
  {pages} {025024} (\bibinfo {year} {2011})},\ \Eprint
  {https://arxiv.org/abs/1010.5118} {arXiv:1010.5118 [gr-qc]} \BibitemShut
  {NoStop}%
\bibitem [{\citenamefont {Wang}\ \emph {et~al.}(2009)\citenamefont {Wang},
  \citenamefont {Pan}, \citenamefont {Chen},\ and\ \citenamefont
  {Jing}}]{Wang:2009wg}%
  \BibitemOpen
  \bibfield  {author} {\bibinfo {author} {\bibfnamefont {J.}~\bibnamefont
  {Wang}}, \bibinfo {author} {\bibfnamefont {Q.}~\bibnamefont {Pan}}, \bibinfo
  {author} {\bibfnamefont {S.}~\bibnamefont {Chen}},\ and\ \bibinfo {author}
  {\bibfnamefont {J.}~\bibnamefont {Jing}},\ }\bibfield  {title} {\bibinfo
  {title} {{Entanglement of coupled massive scalar field in background of
  dilaton black hole}},\ }\href
  {https://doi.org/10.1016/j.physletb.2009.05.028} {\bibfield  {journal}
  {\bibinfo  {journal} {Phys. Lett. B}\ }\textbf {\bibinfo {volume} {677}},\
  \bibinfo {pages} {186} (\bibinfo {year} {2009})},\ \Eprint
  {https://arxiv.org/abs/0905.3226} {arXiv:0905.3226 [gr-qc]} \BibitemShut
  {NoStop}%
\bibitem [{\citenamefont {He}\ \emph {et~al.}(2016)\citenamefont {He},
  \citenamefont {Xu},\ and\ \citenamefont {Ye}}]{He:2016ujs}%
  \BibitemOpen
  \bibfield  {author} {\bibinfo {author} {\bibfnamefont {J.}~\bibnamefont
  {He}}, \bibinfo {author} {\bibfnamefont {S.}~\bibnamefont {Xu}},\ and\
  \bibinfo {author} {\bibfnamefont {L.}~\bibnamefont {Ye}},\ }\bibfield
  {title} {\bibinfo {title} {{Measurement-induced-nonlocality for Dirac
  particles in Garfinkle\textendash{}Horowitz\textendash{}Strominger dilation
  space\textendash{}time}},\ }\href
  {https://doi.org/10.1016/j.physletb.2016.02.073} {\bibfield  {journal}
  {\bibinfo  {journal} {Phys. Lett. B}\ }\textbf {\bibinfo {volume} {756}},\
  \bibinfo {pages} {278} (\bibinfo {year} {2016})}\BibitemShut {NoStop}%
\bibitem [{\citenamefont {Wang}\ \emph {et~al.}(2018)\citenamefont {Wang},
  \citenamefont {Jing},\ and\ \citenamefont {Fan}}]{Wang:2017tfv}%
  \BibitemOpen
  \bibfield  {author} {\bibinfo {author} {\bibfnamefont {J.}~\bibnamefont
  {Wang}}, \bibinfo {author} {\bibfnamefont {J.}~\bibnamefont {Jing}},\ and\
  \bibinfo {author} {\bibfnamefont {H.}~\bibnamefont {Fan}},\ }\bibfield
  {title} {\bibinfo {title} {{Monogamy of Einstein-Podolsky-Rosen Steering in
  the Background of an Asymptotically Flat Black Hole}},\ }\href
  {https://doi.org/10.1002/andp.201700261} {\bibfield  {journal} {\bibinfo
  {journal} {Annalen Phys.}\ }\textbf {\bibinfo {volume} {530}},\ \bibinfo
  {pages} {1700261} (\bibinfo {year} {2018})},\ \Eprint
  {https://arxiv.org/abs/1710.04409} {arXiv:1710.04409 [quant-ph]} \BibitemShut
  {NoStop}%
\bibitem [{\citenamefont {Belfiglio}\ \emph {et~al.}(2021)\citenamefont
  {Belfiglio}, \citenamefont {Luongo},\ and\ \citenamefont
  {Mancini}}]{Belfiglio:2021mnr}%
  \BibitemOpen
  \bibfield  {author} {\bibinfo {author} {\bibfnamefont {A.}~\bibnamefont
  {Belfiglio}}, \bibinfo {author} {\bibfnamefont {O.}~\bibnamefont {Luongo}},\
  and\ \bibinfo {author} {\bibfnamefont {S.}~\bibnamefont {Mancini}},\
  }\bibfield  {title} {\bibinfo {title} {{Entanglement production in
  Einstein-Cartan theory}},\ }\href
  {https://doi.org/10.1103/PhysRevD.104.043523} {\bibfield  {journal} {\bibinfo
   {journal} {Phys. Rev. D}\ }\textbf {\bibinfo {volume} {104}},\ \bibinfo
  {pages} {043523} (\bibinfo {year} {2021})},\ \Eprint
  {https://arxiv.org/abs/2101.11567} {arXiv:2101.11567 [gr-qc]} \BibitemShut
  {NoStop}%
\bibitem [{\citenamefont {Bueley}\ \emph {et~al.}(2022)\citenamefont {Bueley},
  \citenamefont {Huang}, \citenamefont {Gallock-Yoshimura},\ and\ \citenamefont
  {Mann}}]{Bueley:2022ple}%
  \BibitemOpen
  \bibfield  {author} {\bibinfo {author} {\bibfnamefont {K.}~\bibnamefont
  {Bueley}}, \bibinfo {author} {\bibfnamefont {L.}~\bibnamefont {Huang}},
  \bibinfo {author} {\bibfnamefont {K.}~\bibnamefont {Gallock-Yoshimura}},\
  and\ \bibinfo {author} {\bibfnamefont {R.~B.}\ \bibnamefont {Mann}},\
  }\bibfield  {title} {\bibinfo {title} {{Harvesting mutual information from
  BTZ black hole spacetime}},\ }\href
  {https://doi.org/10.1103/PhysRevD.106.025010} {\bibfield  {journal} {\bibinfo
   {journal} {Phys. Rev. D}\ }\textbf {\bibinfo {volume} {106}},\ \bibinfo
  {pages} {025010} (\bibinfo {year} {2022})},\ \Eprint
  {https://arxiv.org/abs/2205.07891} {arXiv:2205.07891 [quant-ph]} \BibitemShut
  {NoStop}%
\bibitem [{\citenamefont {Wu}\ \emph {et~al.}(2023{\natexlab{a}})\citenamefont
  {Wu}, \citenamefont {Fan}, \citenamefont {Wang}, \citenamefont {Wu},
  \citenamefont {Huang},\ and\ \citenamefont {Zeng}}]{Wu:2023sye}%
  \BibitemOpen
  \bibfield  {author} {\bibinfo {author} {\bibfnamefont {S.~M.}\ \bibnamefont
  {Wu}}, \bibinfo {author} {\bibfnamefont {X.~W.}\ \bibnamefont {Fan}},
  \bibinfo {author} {\bibfnamefont {R.~D.}\ \bibnamefont {Wang}}, \bibinfo
  {author} {\bibfnamefont {H.~Y.}\ \bibnamefont {Wu}}, \bibinfo {author}
  {\bibfnamefont {X.~L.}\ \bibnamefont {Huang}},\ and\ \bibinfo {author}
  {\bibfnamefont {H.~S.}\ \bibnamefont {Zeng}},\ }\bibfield  {title} {\bibinfo
  {title} {{Does Hawking effect always degrade fidelity of quantum
  teleportation in Schwarzschild spacetime?}},\ }\href
  {https://doi.org/10.1007/JHEP11(2023)232} {\bibfield  {journal} {\bibinfo
  {journal} {JHEP}\ }\textbf {\bibinfo {volume} {11}},\ \bibinfo {pages}
  {232}},\ \Eprint {https://arxiv.org/abs/2304.00984} {arXiv:2304.00984
  [gr-qc]} \BibitemShut {NoStop}%
\bibitem [{\citenamefont {Wu}\ \emph {et~al.}(2024{\natexlab{a}})\citenamefont
  {Wu}, \citenamefont {Teng}, \citenamefont {Li}, \citenamefont {Li},
  \citenamefont {Liu},\ and\ \citenamefont {Wang}}]{Wu:2023spa}%
  \BibitemOpen
  \bibfield  {author} {\bibinfo {author} {\bibfnamefont {S.~M.}\ \bibnamefont
  {Wu}}, \bibinfo {author} {\bibfnamefont {X.~W.}\ \bibnamefont {Teng}},
  \bibinfo {author} {\bibfnamefont {J.~X.}\ \bibnamefont {Li}}, \bibinfo
  {author} {\bibfnamefont {S.~H.}\ \bibnamefont {Li}}, \bibinfo {author}
  {\bibfnamefont {T.~H.}\ \bibnamefont {Liu}},\ and\ \bibinfo {author}
  {\bibfnamefont {J.}~\bibnamefont {Wang}},\ }\bibfield  {title} {\bibinfo
  {title} {{Genuinely accessible and inaccessible entanglement in Schwarzschild
  black hole}},\ }\href {https://doi.org/10.1016/j.physletb.2023.138334}
  {\bibfield  {journal} {\bibinfo  {journal} {Phys. Lett. B}\ }\textbf
  {\bibinfo {volume} {848}},\ \bibinfo {pages} {138334} (\bibinfo {year}
  {2024}{\natexlab{a}})},\ \Eprint {https://arxiv.org/abs/2311.12362}
  {arXiv:2311.12362 [gr-qc]} \BibitemShut {NoStop}%
\bibitem [{\citenamefont {Mondal}\ and\ \citenamefont
  {Wen}(2022)}]{Mondal:2022aev}%
  \BibitemOpen
  \bibfield  {author} {\bibinfo {author} {\bibfnamefont {S.}~\bibnamefont
  {Mondal}}\ and\ \bibinfo {author} {\bibfnamefont {W.~Y.}\ \bibnamefont
  {Wen}},\ }\bibfield  {title} {\bibinfo {title} {{Entanglement at the
  soft-hair horizon}},\ }\href {https://doi.org/10.1016/j.physletb.2022.137385}
  {\bibfield  {journal} {\bibinfo  {journal} {Phys. Lett. B}\ }\textbf
  {\bibinfo {volume} {833}},\ \bibinfo {pages} {137385} (\bibinfo {year}
  {2022})},\ \Eprint {https://arxiv.org/abs/2206.02390} {arXiv:2206.02390
  [hep-th]} \BibitemShut {NoStop}%
\bibitem [{\citenamefont {Wu}\ \emph {et~al.}(2023{\natexlab{b}})\citenamefont
  {Wu}, \citenamefont {Wang}, \citenamefont {Liu}, \citenamefont {Huang},\ and\
  \citenamefont {Zeng}}]{Wu:2022lmc}%
  \BibitemOpen
  \bibfield  {author} {\bibinfo {author} {\bibfnamefont {S.-M.}\ \bibnamefont
  {Wu}}, \bibinfo {author} {\bibfnamefont {C.-X.}\ \bibnamefont {Wang}},
  \bibinfo {author} {\bibfnamefont {D.-D.}\ \bibnamefont {Liu}}, \bibinfo
  {author} {\bibfnamefont {X.-L.}\ \bibnamefont {Huang}},\ and\ \bibinfo
  {author} {\bibfnamefont {H.-S.}\ \bibnamefont {Zeng}},\ }\bibfield  {title}
  {\bibinfo {title} {{Would quantum coherence be increased by curvature effect
  in de Sitter space?}},\ }\href {https://doi.org/10.1007/JHEP02(2023)115}
  {\bibfield  {journal} {\bibinfo  {journal} {JHEP}\ }\textbf {\bibinfo
  {volume} {02}},\ \bibinfo {pages} {115}},\ \Eprint
  {https://arxiv.org/abs/2207.11721} {arXiv:2207.11721 [gr-qc]} \BibitemShut
  {NoStop}%
\bibitem [{\citenamefont {Babakan}\ \emph {et~al.}(2024)\citenamefont
  {Babakan}, \citenamefont {Kashef},\ and\ \citenamefont
  {Memarzadeh}}]{Babakan:2024abb}%
  \BibitemOpen
  \bibfield  {author} {\bibinfo {author} {\bibfnamefont {M.}~\bibnamefont
  {Babakan}}, \bibinfo {author} {\bibfnamefont {A.}~\bibnamefont {Kashef}},\
  and\ \bibinfo {author} {\bibfnamefont {L.}~\bibnamefont {Memarzadeh}},\
  }\bibfield  {title} {\bibinfo {title} {{Entanglement degradation under local
  dissipative Landau-Zener noise}},\ }\href
  {https://doi.org/10.1103/PhysRevA.110.012456} {\bibfield  {journal} {\bibinfo
   {journal} {Phys. Rev. A}\ }\textbf {\bibinfo {volume} {110}},\ \bibinfo
  {pages} {012456} (\bibinfo {year} {2024})}\BibitemShut {NoStop}%
\bibitem [{\citenamefont {Wu}\ \emph {et~al.}(2024{\natexlab{b}})\citenamefont
  {Wu}, \citenamefont {Wang}, \citenamefont {Huang},\ and\ \citenamefont
  {Wang}}]{Wu:2024qhd}%
  \BibitemOpen
  \bibfield  {author} {\bibinfo {author} {\bibfnamefont {S.-M.}\ \bibnamefont
  {Wu}}, \bibinfo {author} {\bibfnamefont {R.-D.}\ \bibnamefont {Wang}},
  \bibinfo {author} {\bibfnamefont {X.-L.}\ \bibnamefont {Huang}},\ and\
  \bibinfo {author} {\bibfnamefont {Z.}~\bibnamefont {Wang}},\ }\bibfield
  {title} {\bibinfo {title} {{Does gravitational wave assist vacuum steering
  and Bell nonlocality?}},\ }\href {https://doi.org/10.1007/JHEP07(2024)155}
  {\bibfield  {journal} {\bibinfo  {journal} {JHEP}\ }\textbf {\bibinfo
  {volume} {07}},\ \bibinfo {pages} {155}},\ \Eprint
  {https://arxiv.org/abs/2405.07235} {arXiv:2405.07235 [gr-qc]} \BibitemShut
  {NoStop}%
\bibitem [{\citenamefont {Li}\ and\ \citenamefont {Wu}(2024)}]{Wu:2024BF}%
  \BibitemOpen
  \bibfield  {author} {\bibinfo {author} {\bibfnamefont {W.-M.}\ \bibnamefont
  {Li}}\ and\ \bibinfo {author} {\bibfnamefont {S.-M.}\ \bibnamefont {Wu}},\
  }\bibfield  {title} {\bibinfo {title} {{Bosonic and fermionic coherence of
  N-partite states in the background of a dilaton black hole}},\ }\href
  {https://doi.org/10.1007/JHEP09%282024%29144} {\bibfield  {journal} {\bibinfo
   {journal} {JHEP}\ }\textbf {\bibinfo {volume} {09}},\ \bibinfo {pages}
  {144}},\ \Eprint {https://arxiv.org/abs/2407.07688} {arXiv:2407.07688
  [gr-qc]} \BibitemShut {NoStop}%
\bibitem [{\citenamefont {Liu}\ \emph {et~al.}(2024{\natexlab{a}})\citenamefont
  {Liu}, \citenamefont {Liu}, \citenamefont {Wen},\ and\ \citenamefont
  {Wang}}]{Liu:2024pse}%
  \BibitemOpen
  \bibfield  {author} {\bibinfo {author} {\bibfnamefont {Q.}~\bibnamefont
  {Liu}}, \bibinfo {author} {\bibfnamefont {T.}~\bibnamefont {Liu}}, \bibinfo
  {author} {\bibfnamefont {C.}~\bibnamefont {Wen}},\ and\ \bibinfo {author}
  {\bibfnamefont {J.}~\bibnamefont {Wang}},\ }\bibfield  {title} {\bibinfo
  {title} {{Optimal quantum strategy for locating Unruh channels}},\ }\href
  {https://doi.org/10.1103/PhysRevA.110.022428} {\bibfield  {journal} {\bibinfo
   {journal} {Phys. Rev. A}\ }\textbf {\bibinfo {volume} {110}},\ \bibinfo
  {pages} {022428} (\bibinfo {year} {2024}{\natexlab{a}})},\ \Eprint
  {https://arxiv.org/abs/2404.19216} {arXiv:2404.19216 [gr-qc]} \BibitemShut
  {NoStop}%
\bibitem [{\citenamefont {Crispino}\ \emph {et~al.}(2008)\citenamefont
  {Crispino}, \citenamefont {Higuchi},\ and\ \citenamefont
  {Matsas}}]{Crispino:2007eb}%
  \BibitemOpen
  \bibfield  {author} {\bibinfo {author} {\bibfnamefont {L.~C.~B.}\
  \bibnamefont {Crispino}}, \bibinfo {author} {\bibfnamefont {A.}~\bibnamefont
  {Higuchi}},\ and\ \bibinfo {author} {\bibfnamefont {G.~E.~A.}\ \bibnamefont
  {Matsas}},\ }\bibfield  {title} {\bibinfo {title} {{The Unruh effect and its
  applications}},\ }\href {https://doi.org/10.1103/RevModPhys.80.787}
  {\bibfield  {journal} {\bibinfo  {journal} {Rev. Mod. Phys.}\ }\textbf
  {\bibinfo {volume} {80}},\ \bibinfo {pages} {787} (\bibinfo {year} {2008})},\
  \Eprint {https://arxiv.org/abs/0710.5373} {arXiv:0710.5373 [gr-qc]}
  \BibitemShut {NoStop}%
\bibitem [{\citenamefont {Kollas}\ \emph {et~al.}(2023)\citenamefont {Kollas},
  \citenamefont {Moustos},\ and\ \citenamefont {Mu\~noz}}]{Kollas:2022wgj}%
  \BibitemOpen
  \bibfield  {author} {\bibinfo {author} {\bibfnamefont {N.~K.}\ \bibnamefont
  {Kollas}}, \bibinfo {author} {\bibfnamefont {D.}~\bibnamefont {Moustos}},\
  and\ \bibinfo {author} {\bibfnamefont {M.~R.}\ \bibnamefont {Mu\~noz}},\
  }\bibfield  {title} {\bibinfo {title} {{Cohering and decohering power of
  massive scalar fields under instantaneous interactions}},\ }\href
  {https://doi.org/10.1103/PhysRevA.107.022420} {\bibfield  {journal} {\bibinfo
   {journal} {Phys. Rev. A}\ }\textbf {\bibinfo {volume} {107}},\ \bibinfo
  {pages} {022420} (\bibinfo {year} {2023})},\ \Eprint
  {https://arxiv.org/abs/2206.11816} {arXiv:2206.11816 [quant-ph]} \BibitemShut
  {NoStop}%
\bibitem [{\citenamefont {Kostelecky}\ and\ \citenamefont
  {Samuel}(1989)}]{Kostelecky19891}%
  \BibitemOpen
  \bibfield  {author} {\bibinfo {author} {\bibfnamefont {V.~A.}\ \bibnamefont
  {Kostelecky}}\ and\ \bibinfo {author} {\bibfnamefont {S.}~\bibnamefont
  {Samuel}},\ }\bibfield  {title} {\bibinfo {title} {{Phenomenological
  Gravitational Constraints on Strings and Higher Dimensional Theories}},\
  }\href {https://doi.org/10.1103/PhysRevLett.63.224} {\bibfield  {journal}
  {\bibinfo  {journal} {Phys. Rev. Lett.}\ }\textbf {\bibinfo {volume} {63}},\
  \bibinfo {pages} {224} (\bibinfo {year} {1989})}\BibitemShut {NoStop}%
\bibitem [{\citenamefont {Carroll}\ \emph {et~al.}(2001)\citenamefont
  {Carroll}, \citenamefont {Harvey}, \citenamefont {Kostelecky}, \citenamefont
  {Lane},\ and\ \citenamefont {Okamoto}}]{Kostelecky2001}%
  \BibitemOpen
  \bibfield  {author} {\bibinfo {author} {\bibfnamefont {S.~M.}\ \bibnamefont
  {Carroll}}, \bibinfo {author} {\bibfnamefont {J.~A.}\ \bibnamefont {Harvey}},
  \bibinfo {author} {\bibfnamefont {V.~A.}\ \bibnamefont {Kostelecky}},
  \bibinfo {author} {\bibfnamefont {C.~D.}\ \bibnamefont {Lane}},\ and\
  \bibinfo {author} {\bibfnamefont {T.}~\bibnamefont {Okamoto}},\ }\bibfield
  {title} {\bibinfo {title} {{Noncommutative field theory and Lorentz
  violation}},\ }\href {https://doi.org/10.1103/PhysRevLett.87.141601}
  {\bibfield  {journal} {\bibinfo  {journal} {Phys. Rev. Lett.}\ }\textbf
  {\bibinfo {volume} {87}},\ \bibinfo {pages} {141601} (\bibinfo {year}
  {2001})},\ \Eprint {https://arxiv.org/abs/0105082} {arXiv:0105082 [hep-th]}
  \BibitemShut {NoStop}%
\bibitem [{\citenamefont {Choi}\ and\ \citenamefont
  {Park}(2015)}]{Choi:2015bga}%
  \BibitemOpen
  \bibfield  {author} {\bibinfo {author} {\bibfnamefont {K.-S.}\ \bibnamefont
  {Choi}}\ and\ \bibinfo {author} {\bibfnamefont {J.-H.}\ \bibnamefont
  {Park}},\ }\bibfield  {title} {\bibinfo {title} {{Standard Model as a Double
  Field Theory}},\ }\href {https://doi.org/10.1103/PhysRevLett.115.171603}
  {\bibfield  {journal} {\bibinfo  {journal} {Phys. Rev. Lett.}\ }\textbf
  {\bibinfo {volume} {115}},\ \bibinfo {pages} {171603} (\bibinfo {year}
  {2015})},\ \Eprint {https://arxiv.org/abs/1506.05277} {arXiv:1506.05277
  [hep-th]} \BibitemShut {NoStop}%
\bibitem [{\citenamefont {Del~Porro}\ \emph {et~al.}(2023)\citenamefont
  {Del~Porro}, \citenamefont {Herrero-Valea}, \citenamefont {Liberati},\ and\
  \citenamefont {Schneider}}]{DelPorro:2023lbv}%
  \BibitemOpen
  \bibfield  {author} {\bibinfo {author} {\bibfnamefont {F.}~\bibnamefont
  {Del~Porro}}, \bibinfo {author} {\bibfnamefont {M.}~\bibnamefont
  {Herrero-Valea}}, \bibinfo {author} {\bibfnamefont {S.}~\bibnamefont
  {Liberati}},\ and\ \bibinfo {author} {\bibfnamefont {M.}~\bibnamefont
  {Schneider}},\ }\bibfield  {title} {\bibinfo {title} {{Hawking radiation in
  Lorentz violating gravity: a tale of two horizons}},\ }\href
  {https://doi.org/10.1007/JHEP12(2023)094} {\bibfield  {journal} {\bibinfo
  {journal} {JHEP}\ }\textbf {\bibinfo {volume} {12}},\ \bibinfo {pages}
  {094}},\ \Eprint {https://arxiv.org/abs/2310.01472} {arXiv:2310.01472
  [gr-qc]} \BibitemShut {NoStop}%
\bibitem [{\citenamefont {Tian}\ \emph {et~al.}(2022)\citenamefont {Tian},
  \citenamefont {Wu}, \citenamefont {Zhang}, \citenamefont {Jing},\ and\
  \citenamefont {Du}}]{Tian:2022gfa}%
  \BibitemOpen
  \bibfield  {author} {\bibinfo {author} {\bibfnamefont {Z.}~\bibnamefont
  {Tian}}, \bibinfo {author} {\bibfnamefont {L.}~\bibnamefont {Wu}}, \bibinfo
  {author} {\bibfnamefont {L.}~\bibnamefont {Zhang}}, \bibinfo {author}
  {\bibfnamefont {J.}~\bibnamefont {Jing}},\ and\ \bibinfo {author}
  {\bibfnamefont {J.}~\bibnamefont {Du}},\ }\bibfield  {title} {\bibinfo
  {title} {{Probing Lorentz-invariance-violation-induced nonthermal Unruh
  effect in quasi-two-dimensional dipolar condensates}},\ }\href
  {https://doi.org/10.1103/PhysRevD.106.L061701} {\bibfield  {journal}
  {\bibinfo  {journal} {Phys. Rev. D}\ }\textbf {\bibinfo {volume} {106}},\
  \bibinfo {pages} {L061701} (\bibinfo {year} {2022})},\ \Eprint
  {https://arxiv.org/abs/2205.08669} {arXiv:2205.08669 [quant-ph]} \BibitemShut
  {NoStop}%
\bibitem [{\citenamefont {Agarwalla}\ \emph {et~al.}(2023)\citenamefont
  {Agarwalla}, \citenamefont {Das}, \citenamefont {Sahoo},\ and\ \citenamefont
  {Swain}}]{Agarwalla:2023wft}%
  \BibitemOpen
  \bibfield  {author} {\bibinfo {author} {\bibfnamefont {S.~K.}\ \bibnamefont
  {Agarwalla}}, \bibinfo {author} {\bibfnamefont {S.}~\bibnamefont {Das}},
  \bibinfo {author} {\bibfnamefont {S.}~\bibnamefont {Sahoo}},\ and\ \bibinfo
  {author} {\bibfnamefont {P.}~\bibnamefont {Swain}},\ }\bibfield  {title}
  {\bibinfo {title} {{Constraining Lorentz invariance violation with
  next-generation long-baseline experiments}},\ }\href
  {https://doi.org/10.1007/JHEP07(2023)216} {\bibfield  {journal} {\bibinfo
  {journal} {JHEP}\ }\textbf {\bibinfo {volume} {07}},\ \bibinfo {pages}
  {216}},\ \Eprint {https://arxiv.org/abs/2302.12005} {arXiv:2302.12005
  [hep-ph]} \BibitemShut {NoStop}%
\bibitem [{\citenamefont {Jiang}\ \emph {et~al.}(2024)\citenamefont {Jiang},
  \citenamefont {Pecjak}, \citenamefont {Perez},\ and\ \citenamefont
  {Sankaranarayanan}}]{Jiang:2024agx}%
  \BibitemOpen
  \bibfield  {author} {\bibinfo {author} {\bibfnamefont {M.}~\bibnamefont
  {Jiang}}, \bibinfo {author} {\bibfnamefont {B.~D.}\ \bibnamefont {Pecjak}},
  \bibinfo {author} {\bibfnamefont {G.}~\bibnamefont {Perez}},\ and\ \bibinfo
  {author} {\bibfnamefont {S.}~\bibnamefont {Sankaranarayanan}},\ }\bibfield
  {title} {\bibinfo {title} {{Lorentz violating backgrounds from quadratic,
  shift-symmetric, ultralight dark matter}},\ }\href
  {https://doi.org/10.1007/JHEP08(2024)114} {\bibfield  {journal} {\bibinfo
  {journal} {JHEP}\ }\textbf {\bibinfo {volume} {08}},\ \bibinfo {pages}
  {114}},\ \Eprint {https://arxiv.org/abs/2404.17636} {arXiv:2404.17636
  [hep-ph]} \BibitemShut {NoStop}%
\bibitem [{\citenamefont {Kostelecky}(2004)}]{Kostelecky:2003fs}%
  \BibitemOpen
  \bibfield  {author} {\bibinfo {author} {\bibfnamefont {V.~A.}\ \bibnamefont
  {Kostelecky}},\ }\bibfield  {title} {\bibinfo {title} {{Gravity, Lorentz
  violation, and the standard model}},\ }\href
  {https://doi.org/10.1103/PhysRevD.69.105009} {\bibfield  {journal} {\bibinfo
  {journal} {Phys. Rev. D}\ }\textbf {\bibinfo {volume} {69}},\ \bibinfo
  {pages} {105009} (\bibinfo {year} {2004})},\ \Eprint
  {https://arxiv.org/abs/hep-th/0312310} {arXiv:hep-th/0312310} \BibitemShut
  {NoStop}%
\bibitem [{\citenamefont {Bluhm}\ and\ \citenamefont
  {Kostelecky}(2005)}]{Bluhm:2004ep}%
  \BibitemOpen
  \bibfield  {author} {\bibinfo {author} {\bibfnamefont {R.}~\bibnamefont
  {Bluhm}}\ and\ \bibinfo {author} {\bibfnamefont {V.~A.}\ \bibnamefont
  {Kostelecky}},\ }\bibfield  {title} {\bibinfo {title} {{Spontaneous Lorentz
  violation, Nambu-Goldstone modes, and gravity}},\ }\href
  {https://doi.org/10.1103/PhysRevD.71.065008} {\bibfield  {journal} {\bibinfo
  {journal} {Phys. Rev. D}\ }\textbf {\bibinfo {volume} {71}},\ \bibinfo
  {pages} {065008} (\bibinfo {year} {2005})},\ \Eprint
  {https://arxiv.org/abs/hep-th/0412320} {arXiv:hep-th/0412320} \BibitemShut
  {NoStop}%
\bibitem [{\citenamefont {Casana}\ \emph {et~al.}(2018)\citenamefont {Casana},
  \citenamefont {Cavalcante}, \citenamefont {Poulis},\ and\ \citenamefont
  {Santos}}]{Casana2018}%
  \BibitemOpen
  \bibfield  {author} {\bibinfo {author} {\bibfnamefont {R.}~\bibnamefont
  {Casana}}, \bibinfo {author} {\bibfnamefont {A.}~\bibnamefont {Cavalcante}},
  \bibinfo {author} {\bibfnamefont {F.~P.}\ \bibnamefont {Poulis}},\ and\
  \bibinfo {author} {\bibfnamefont {E.~B.}\ \bibnamefont {Santos}},\ }\bibfield
   {title} {\bibinfo {title} {{Exact Schwarzschild-like solution in a bumblebee
  gravity model}},\ }\href {https://doi.org/10.1103/PhysRevD.97.104001}
  {\bibfield  {journal} {\bibinfo  {journal} {Phys. Rev. D}\ }\textbf {\bibinfo
  {volume} {97}},\ \bibinfo {pages} {104001} (\bibinfo {year} {2018})},\
  \Eprint {https://arxiv.org/abs/1711.02273} {arXiv:1711.02273 [gr-qc]}
  \BibitemShut {NoStop}%
\bibitem [{\citenamefont {\"Ovg\"un}\ \emph {et~al.}(2019)\citenamefont
  {\"Ovg\"un}, \citenamefont {Jusufi},\ and\ \citenamefont
  {Sakall\i{}}}]{Ovgun2019}%
  \BibitemOpen
  \bibfield  {author} {\bibinfo {author} {\bibfnamefont {A.}~\bibnamefont
  {\"Ovg\"un}}, \bibinfo {author} {\bibfnamefont {K.}~\bibnamefont {Jusufi}},\
  and\ \bibinfo {author} {\bibfnamefont {I.}~\bibnamefont {Sakall\i{}}},\
  }\bibfield  {title} {\bibinfo {title} {{Exact traversable wormhole solution
  in bumblebee gravity}},\ }\href {https://doi.org/10.1103/PhysRevD.99.024042}
  {\bibfield  {journal} {\bibinfo  {journal} {Phys. Rev. D}\ }\textbf {\bibinfo
  {volume} {99}},\ \bibinfo {pages} {024042} (\bibinfo {year} {2019})},\
  \Eprint {https://arxiv.org/abs/1804.09911} {arXiv:1804.09911 [gr-qc]}
  \BibitemShut {NoStop}%
\bibitem [{\citenamefont {G\"ull\"u}\ and\ \citenamefont
  {\"Ovg\"un}(2022)}]{Gullu2020}%
  \BibitemOpen
  \bibfield  {author} {\bibinfo {author} {\bibfnamefont {I.}~\bibnamefont
  {G\"ull\"u}}\ and\ \bibinfo {author} {\bibfnamefont {A.}~\bibnamefont
  {\"Ovg\"un}},\ }\bibfield  {title} {\bibinfo {title} {{Schwarzschild-like
  black hole with a topological defect in bumblebee gravity}},\ }\href
  {https://doi.org/10.1016/j.aop.2021.168721} {\bibfield  {journal} {\bibinfo
  {journal} {Annals Phys.}\ }\textbf {\bibinfo {volume} {436}},\ \bibinfo
  {pages} {168721} (\bibinfo {year} {2022})},\ \Eprint
  {https://arxiv.org/abs/2012.02611} {arXiv:2012.02611 [gr-qc]} \BibitemShut
  {NoStop}%
\bibitem [{\citenamefont {Poulis}\ and\ \citenamefont
  {Soares}(2022)}]{Poulis:2021nqh}%
  \BibitemOpen
  \bibfield  {author} {\bibinfo {author} {\bibfnamefont {F.~P.}\ \bibnamefont
  {Poulis}}\ and\ \bibinfo {author} {\bibfnamefont {M.~A.~C.}\ \bibnamefont
  {Soares}},\ }\bibfield  {title} {\bibinfo {title} {{Exact modifications on a
  vacuum spacetime due to a gradient bumblebee field at its vacuum expectation
  value}},\ }\href {https://doi.org/10.1140/epjc/s10052-022-10547-y} {\bibfield
   {journal} {\bibinfo  {journal} {Eur. Phys. J. C}\ }\textbf {\bibinfo
  {volume} {82}},\ \bibinfo {pages} {613} (\bibinfo {year} {2022})},\ \Eprint
  {https://arxiv.org/abs/2112.04040} {arXiv:2112.04040 [gr-qc]} \BibitemShut
  {NoStop}%
\bibitem [{\citenamefont {Maluf}\ and\ \citenamefont
  {Neves}(2021)}]{Maluf2021}%
  \BibitemOpen
  \bibfield  {author} {\bibinfo {author} {\bibfnamefont {R.~V.}\ \bibnamefont
  {Maluf}}\ and\ \bibinfo {author} {\bibfnamefont {J.~C.~S.}\ \bibnamefont
  {Neves}},\ }\bibfield  {title} {\bibinfo {title} {{Black holes with a
  cosmological constant in bumblebee gravity}},\ }\href
  {https://doi.org/10.1103/PhysRevD.103.044002} {\bibfield  {journal} {\bibinfo
   {journal} {Phys. Rev. D}\ }\textbf {\bibinfo {volume} {103}},\ \bibinfo
  {pages} {044002} (\bibinfo {year} {2021})},\ \Eprint
  {https://arxiv.org/abs/2011.12841} {arXiv:2011.12841 [gr-qc]} \BibitemShut
  {NoStop}%
\bibitem [{\citenamefont {Xu}\ \emph {et~al.}(2023{\natexlab{a}})\citenamefont
  {Xu}, \citenamefont {Liang},\ and\ \citenamefont {Shao}}]{Xu:2022frb}%
  \BibitemOpen
  \bibfield  {author} {\bibinfo {author} {\bibfnamefont {R.}~\bibnamefont
  {Xu}}, \bibinfo {author} {\bibfnamefont {D.}~\bibnamefont {Liang}},\ and\
  \bibinfo {author} {\bibfnamefont {L.}~\bibnamefont {Shao}},\ }\bibfield
  {title} {\bibinfo {title} {{Static spherical vacuum solutions in the
  bumblebee gravity model}},\ }\href
  {https://doi.org/10.1103/PhysRevD.107.024011} {\bibfield  {journal} {\bibinfo
   {journal} {Phys. Rev. D}\ }\textbf {\bibinfo {volume} {107}},\ \bibinfo
  {pages} {024011} (\bibinfo {year} {2023}{\natexlab{a}})},\ \Eprint
  {https://arxiv.org/abs/2209.02209} {arXiv:2209.02209 [gr-qc]} \BibitemShut
  {NoStop}%
\bibitem [{\citenamefont {Ding}\ \emph {et~al.}(2022)\citenamefont {Ding},
  \citenamefont {Chen},\ and\ \citenamefont {Fu}}]{Ding2022}%
  \BibitemOpen
  \bibfield  {author} {\bibinfo {author} {\bibfnamefont {C.}~\bibnamefont
  {Ding}}, \bibinfo {author} {\bibfnamefont {X.}~\bibnamefont {Chen}},\ and\
  \bibinfo {author} {\bibfnamefont {X.}~\bibnamefont {Fu}},\ }\bibfield
  {title} {\bibinfo {title} {{Einstein-Gauss-Bonnet gravity coupled to
  bumblebee field in four dimensional spacetime}},\ }\href
  {https://doi.org/10.1016/j.nuclphysb.2022.115688} {\bibfield  {journal}
  {\bibinfo  {journal} {Nucl. Phys. B}\ }\textbf {\bibinfo {volume} {975}},\
  \bibinfo {pages} {115688} (\bibinfo {year} {2022})},\ \Eprint
  {https://arxiv.org/abs/2102.13335} {arXiv:2102.13335 [gr-qc]} \BibitemShut
  {NoStop}%
\bibitem [{\citenamefont {Liu}\ \emph {et~al.}(2023{\natexlab{a}})\citenamefont
  {Liu}, \citenamefont {Fang}, \citenamefont {Jing},\ and\ \citenamefont
  {Wang}}]{Liu:2022dcn}%
  \BibitemOpen
  \bibfield  {author} {\bibinfo {author} {\bibfnamefont {W.}~\bibnamefont
  {Liu}}, \bibinfo {author} {\bibfnamefont {X.}~\bibnamefont {Fang}}, \bibinfo
  {author} {\bibfnamefont {J.}~\bibnamefont {Jing}},\ and\ \bibinfo {author}
  {\bibfnamefont {J.}~\bibnamefont {Wang}},\ }\bibfield  {title} {\bibinfo
  {title} {{QNMs of slowly rotating Einstein\textendash{}Bumblebee black
  hole}},\ }\href {https://doi.org/10.1140/epjc/s10052-023-11231-5} {\bibfield
  {journal} {\bibinfo  {journal} {Eur. Phys. J. C}\ }\textbf {\bibinfo {volume}
  {83}},\ \bibinfo {pages} {83} (\bibinfo {year} {2023}{\natexlab{a}})},\
  \Eprint {https://arxiv.org/abs/2211.03156} {arXiv:2211.03156 [gr-qc]}
  \BibitemShut {NoStop}%
\bibitem [{\citenamefont {Mai}\ \emph {et~al.}(2023)\citenamefont {Mai},
  \citenamefont {Xu}, \citenamefont {Liang},\ and\ \citenamefont
  {Shao}}]{Mai:2023ggs}%
  \BibitemOpen
  \bibfield  {author} {\bibinfo {author} {\bibfnamefont {Z.-F.}\ \bibnamefont
  {Mai}}, \bibinfo {author} {\bibfnamefont {R.}~\bibnamefont {Xu}}, \bibinfo
  {author} {\bibfnamefont {D.}~\bibnamefont {Liang}},\ and\ \bibinfo {author}
  {\bibfnamefont {L.}~\bibnamefont {Shao}},\ }\bibfield  {title} {\bibinfo
  {title} {{Extended thermodynamics of the bumblebee black holes}},\ }\href
  {https://doi.org/10.1103/PhysRevD.108.024004} {\bibfield  {journal} {\bibinfo
   {journal} {Phys. Rev. D}\ }\textbf {\bibinfo {volume} {108}},\ \bibinfo
  {pages} {024004} (\bibinfo {year} {2023})},\ \Eprint
  {https://arxiv.org/abs/2304.08030} {arXiv:2304.08030 [gr-qc]} \BibitemShut
  {NoStop}%
\bibitem [{\citenamefont {Xu}\ \emph {et~al.}(2023{\natexlab{b}})\citenamefont
  {Xu}, \citenamefont {Liang},\ and\ \citenamefont {Shao}}]{Xu:2023xqh}%
  \BibitemOpen
  \bibfield  {author} {\bibinfo {author} {\bibfnamefont {R.}~\bibnamefont
  {Xu}}, \bibinfo {author} {\bibfnamefont {D.}~\bibnamefont {Liang}},\ and\
  \bibinfo {author} {\bibfnamefont {L.}~\bibnamefont {Shao}},\ }\bibfield
  {title} {\bibinfo {title} {{Bumblebee Black Holes in Light of Event Horizon
  Telescope Observations}},\ }\href {https://doi.org/10.3847/1538-4357/acbdfb}
  {\bibfield  {journal} {\bibinfo  {journal} {Astrophys. J.}\ }\textbf
  {\bibinfo {volume} {945}},\ \bibinfo {pages} {148} (\bibinfo {year}
  {2023}{\natexlab{b}})},\ \Eprint {https://arxiv.org/abs/2302.05671}
  {arXiv:2302.05671 [gr-qc]} \BibitemShut {NoStop}%
\bibitem [{\citenamefont {Zhang}\ \emph {et~al.}(2023)\citenamefont {Zhang},
  \citenamefont {Wang},\ and\ \citenamefont {Jing}}]{Zhang:2023wwk}%
  \BibitemOpen
  \bibfield  {author} {\bibinfo {author} {\bibfnamefont {X.}~\bibnamefont
  {Zhang}}, \bibinfo {author} {\bibfnamefont {M.}~\bibnamefont {Wang}},\ and\
  \bibinfo {author} {\bibfnamefont {J.}~\bibnamefont {Jing}},\ }\bibfield
  {title} {\bibinfo {title} {{Quasinormal modes and late time tails of
  perturbation fields on a Schwarzschild-like black hole with a global monopole
  in the Einstein-bumblebee theory}},\ }\href
  {https://doi.org/10.1007/s11433-023-2153-6} {\bibfield  {journal} {\bibinfo
  {journal} {Sci. China Phys. Mech. Astron.}\ }\textbf {\bibinfo {volume}
  {66}},\ \bibinfo {pages} {100411} (\bibinfo {year} {2023})},\ \Eprint
  {https://arxiv.org/abs/2307.10856} {arXiv:2307.10856 [gr-qc]} \BibitemShut
  {NoStop}%
\bibitem [{\citenamefont {Lin}\ \emph {et~al.}(2023)\citenamefont {Lin},
  \citenamefont {Jiang},\ and\ \citenamefont {Zhai}}]{Lin:2023foj}%
  \BibitemOpen
  \bibfield  {author} {\bibinfo {author} {\bibfnamefont {R.-H.}\ \bibnamefont
  {Lin}}, \bibinfo {author} {\bibfnamefont {R.}~\bibnamefont {Jiang}},\ and\
  \bibinfo {author} {\bibfnamefont {X.-H.}\ \bibnamefont {Zhai}},\ }\bibfield
  {title} {\bibinfo {title} {{Quasinormal modes of the spherical bumblebee
  black holes with a global monopole}},\ }\href
  {https://doi.org/10.1140/epjc/s10052-023-11899-9} {\bibfield  {journal}
  {\bibinfo  {journal} {Eur. Phys. J. C}\ }\textbf {\bibinfo {volume} {83}},\
  \bibinfo {pages} {720} (\bibinfo {year} {2023})},\ \Eprint
  {https://arxiv.org/abs/2308.01575} {arXiv:2308.01575 [gr-qc]} \BibitemShut
  {NoStop}%
\bibitem [{\citenamefont {Chen}\ \emph {et~al.}(2023)\citenamefont {Chen},
  \citenamefont {Pan},\ and\ \citenamefont {Jing}}]{Chen:2023cjd}%
  \BibitemOpen
  \bibfield  {author} {\bibinfo {author} {\bibfnamefont {C.}~\bibnamefont
  {Chen}}, \bibinfo {author} {\bibfnamefont {Q.}~\bibnamefont {Pan}},\ and\
  \bibinfo {author} {\bibfnamefont {J.}~\bibnamefont {Jing}},\ }\bibfield
  {title} {\bibinfo {title} {{Quasinormal modes of a scalar perturbation around
  a rotating BTZ-like black hole in Einstein-bumblebee gravity}},\ }\href
  {https://doi.org/10.1016/j.physletb.2023.138186} {\bibfield  {journal}
  {\bibinfo  {journal} {Phys. Lett. B}\ }\textbf {\bibinfo {volume} {846}},\
  \bibinfo {pages} {138186} (\bibinfo {year} {2023})},\ \Eprint
  {https://arxiv.org/abs/2302.05861} {arXiv:2302.05861 [gr-qc]} \BibitemShut
  {NoStop}%
\bibitem [{\citenamefont {Chen}\ \emph {et~al.}(2020)\citenamefont {Chen},
  \citenamefont {Wang},\ and\ \citenamefont {Jing}}]{Chen2020}%
  \BibitemOpen
  \bibfield  {author} {\bibinfo {author} {\bibfnamefont {S.}~\bibnamefont
  {Chen}}, \bibinfo {author} {\bibfnamefont {M.}~\bibnamefont {Wang}},\ and\
  \bibinfo {author} {\bibfnamefont {J.}~\bibnamefont {Jing}},\ }\bibfield
  {title} {\bibinfo {title} {{Polarization effects in Kerr black hole shadow
  due to the coupling between photon and bumblebee field}},\ }\href
  {https://doi.org/10.1007/JHEP07(2020)054} {\bibfield  {journal} {\bibinfo
  {journal} {JHEP}\ }\textbf {\bibinfo {volume} {07}},\ \bibinfo {pages}
  {054}},\ \Eprint {https://arxiv.org/abs/2004.08857} {arXiv:2004.08857
  [gr-qc]} \BibitemShut {NoStop}%
\bibitem [{\citenamefont {Wang}\ \emph {et~al.}(2022)\citenamefont {Wang},
  \citenamefont {Chen},\ and\ \citenamefont {Jing}}]{Wang:2021gtd}%
  \BibitemOpen
  \bibfield  {author} {\bibinfo {author} {\bibfnamefont {Z.}~\bibnamefont
  {Wang}}, \bibinfo {author} {\bibfnamefont {S.}~\bibnamefont {Chen}},\ and\
  \bibinfo {author} {\bibfnamefont {J.}~\bibnamefont {Jing}},\ }\bibfield
  {title} {\bibinfo {title} {{Constraint on parameters of a rotating black hole
  in Einstein-bumblebee theory by quasi-periodic oscillations}},\ }\href
  {https://doi.org/10.1140/epjc/s10052-022-10475-x} {\bibfield  {journal}
  {\bibinfo  {journal} {Eur. Phys. J. C}\ }\textbf {\bibinfo {volume} {82}},\
  \bibinfo {pages} {528} (\bibinfo {year} {2022})},\ \Eprint
  {https://arxiv.org/abs/2112.02895} {arXiv:2112.02895 [gr-qc]} \BibitemShut
  {NoStop}%
\bibitem [{\citenamefont {Liu}\ \emph {et~al.}(2024{\natexlab{b}})\citenamefont
  {Liu}, \citenamefont {Fang}, \citenamefont {Jing},\ and\ \citenamefont
  {Wang}}]{Liu:2024oeq}%
  \BibitemOpen
  \bibfield  {author} {\bibinfo {author} {\bibfnamefont {W.}~\bibnamefont
  {Liu}}, \bibinfo {author} {\bibfnamefont {X.}~\bibnamefont {Fang}}, \bibinfo
  {author} {\bibfnamefont {J.}~\bibnamefont {Jing}},\ and\ \bibinfo {author}
  {\bibfnamefont {J.}~\bibnamefont {Wang}},\ }\bibfield  {title} {\bibinfo
  {title} {{Lorentz violation induces isospectrality breaking in
  Einstein-bumblebee gravity theory}},\ }\href
  {https://doi.org/10.1007/s11433-024-2405-y} {\bibfield  {journal} {\bibinfo
  {journal} {Sci. China Phys. Mech. Astron.}\ }\textbf {\bibinfo {volume}
  {67}},\ \bibinfo {pages} {280413} (\bibinfo {year} {2024}{\natexlab{b}})},\
  \Eprint {https://arxiv.org/abs/2402.09686} {arXiv:2402.09686 [gr-qc]}
  \BibitemShut {NoStop}%
\bibitem [{\citenamefont {Mai}\ \emph {et~al.}(2024)\citenamefont {Mai},
  \citenamefont {Xu}, \citenamefont {Liang},\ and\ \citenamefont
  {Shao}}]{Mai:2024lgk}%
  \BibitemOpen
  \bibfield  {author} {\bibinfo {author} {\bibfnamefont {Z.-F.}\ \bibnamefont
  {Mai}}, \bibinfo {author} {\bibfnamefont {R.}~\bibnamefont {Xu}}, \bibinfo
  {author} {\bibfnamefont {D.}~\bibnamefont {Liang}},\ and\ \bibinfo {author}
  {\bibfnamefont {L.}~\bibnamefont {Shao}},\ }\bibfield  {title} {\bibinfo
  {title} {{Dynamic instability analysis for bumblebee black holes: The odd
  parity}},\ }\href {https://doi.org/10.1103/PhysRevD.109.084076} {\bibfield
  {journal} {\bibinfo  {journal} {Phys. Rev. D}\ }\textbf {\bibinfo {volume}
  {109}},\ \bibinfo {pages} {084076} (\bibinfo {year} {2024})},\ \Eprint
  {https://arxiv.org/abs/2401.07757} {arXiv:2401.07757 [gr-qc]} \BibitemShut
  {NoStop}%
\bibitem [{\citenamefont {Liang}\ \emph {et~al.}(2023)\citenamefont {Liang},
  \citenamefont {Xu}, \citenamefont {Mai},\ and\ \citenamefont
  {Shao}}]{Liang:2022gdk}%
  \BibitemOpen
  \bibfield  {author} {\bibinfo {author} {\bibfnamefont {D.}~\bibnamefont
  {Liang}}, \bibinfo {author} {\bibfnamefont {R.}~\bibnamefont {Xu}}, \bibinfo
  {author} {\bibfnamefont {Z.-F.}\ \bibnamefont {Mai}},\ and\ \bibinfo {author}
  {\bibfnamefont {L.}~\bibnamefont {Shao}},\ }\bibfield  {title} {\bibinfo
  {title} {{Probing vector hair of black holes with extreme-mass-ratio
  inspirals}},\ }\href {https://doi.org/10.1103/PhysRevD.107.044053} {\bibfield
   {journal} {\bibinfo  {journal} {Phys. Rev. D}\ }\textbf {\bibinfo {volume}
  {107}},\ \bibinfo {pages} {044053} (\bibinfo {year} {2023})},\ \Eprint
  {https://arxiv.org/abs/2212.09346} {arXiv:2212.09346 [gr-qc]} \BibitemShut
  {NoStop}%
\bibitem [{\citenamefont {Wald}(1993)}]{Wald:1993nt}%
  \BibitemOpen
  \bibfield  {author} {\bibinfo {author} {\bibfnamefont {R.~M.}\ \bibnamefont
  {Wald}},\ }\bibfield  {title} {\bibinfo {title} {{Black hole entropy is the
  Noether charge}},\ }\href {https://doi.org/10.1103/PhysRevD.48.R3427}
  {\bibfield  {journal} {\bibinfo  {journal} {Phys. Rev. D}\ }\textbf {\bibinfo
  {volume} {48}},\ \bibinfo {pages} {R3427} (\bibinfo {year} {1993})},\ \Eprint
  {https://arxiv.org/abs/9307038} {arXiv:9307038 [gr-qc]} \BibitemShut
  {NoStop}%
\bibitem [{\citenamefont {Kibble}(1976)}]{Kibble:1976sj}%
  \BibitemOpen
  \bibfield  {author} {\bibinfo {author} {\bibfnamefont {T.~W.~B.}\
  \bibnamefont {Kibble}},\ }\bibfield  {title} {\bibinfo {title} {{Topology of
  Cosmic Domains and Strings}},\ }\href
  {https://doi.org/10.1088/0305-4470/9/8/029} {\bibfield  {journal} {\bibinfo
  {journal} {J. Phys. A}\ }\textbf {\bibinfo {volume} {9}},\ \bibinfo {pages}
  {1387} (\bibinfo {year} {1976})}\BibitemShut {NoStop}%
\bibitem [{\citenamefont {Vilenkin}(1985)}]{Vilenkin:1984ib}%
  \BibitemOpen
  \bibfield  {author} {\bibinfo {author} {\bibfnamefont {A.}~\bibnamefont
  {Vilenkin}},\ }\bibfield  {title} {\bibinfo {title} {{Cosmic Strings and
  Domain Walls}},\ }\href {https://doi.org/10.1016/0370-1573(85)90033-X}
  {\bibfield  {journal} {\bibinfo  {journal} {Phys. Rept.}\ }\textbf {\bibinfo
  {volume} {121}},\ \bibinfo {pages} {263} (\bibinfo {year}
  {1985})}\BibitemShut {NoStop}%
\bibitem [{\citenamefont {Sen}\ \emph {et~al.}(2024)\citenamefont {Sen},
  \citenamefont {Mukherjee},\ and\ \citenamefont {Gangopadhyay}}]{Sen:2023sfb}%
  \BibitemOpen
  \bibfield  {author} {\bibinfo {author} {\bibfnamefont {S.}~\bibnamefont
  {Sen}}, \bibinfo {author} {\bibfnamefont {A.}~\bibnamefont {Mukherjee}},\
  and\ \bibinfo {author} {\bibfnamefont {S.}~\bibnamefont {Gangopadhyay}},\
  }\bibfield  {title} {\bibinfo {title} {{Entanglement degradation as a tool to
  detect signatures of modified gravity}},\ }\href
  {https://doi.org/10.1103/PhysRevD.109.046012} {\bibfield  {journal} {\bibinfo
   {journal} {Phys. Rev. D}\ }\textbf {\bibinfo {volume} {109}},\ \bibinfo
  {pages} {046012} (\bibinfo {year} {2024})},\ \Eprint
  {https://arxiv.org/abs/2308.04925} {arXiv:2308.04925 [hep-th]} \BibitemShut
  {NoStop}%
\bibitem [{\citenamefont {Pan}\ and\ \citenamefont {Jing}(2008)}]{Pan:2008yr}%
  \BibitemOpen
  \bibfield  {author} {\bibinfo {author} {\bibfnamefont {Q.}~\bibnamefont
  {Pan}}\ and\ \bibinfo {author} {\bibfnamefont {J.}~\bibnamefont {Jing}},\
  }\bibfield  {title} {\bibinfo {title} {{Hawking radiation, Entanglement and
  Teleportation in background of an asymptotically flat static black hole}},\
  }\href {https://doi.org/10.1103/PhysRevD.78.065015} {\bibfield  {journal}
  {\bibinfo  {journal} {Phys. Rev. D}\ }\textbf {\bibinfo {volume} {78}},\
  \bibinfo {pages} {065015} (\bibinfo {year} {2008})},\ \Eprint
  {https://arxiv.org/abs/0809.0811} {arXiv:0809.0811 [gr-qc]} \BibitemShut
  {NoStop}%
\bibitem [{\citenamefont {Wu}\ and\ \citenamefont {Zeng}(2022)}]{Wu:2022xwy}%
  \BibitemOpen
  \bibfield  {author} {\bibinfo {author} {\bibfnamefont {S.~M.}\ \bibnamefont
  {Wu}}\ and\ \bibinfo {author} {\bibfnamefont {H.~S.}\ \bibnamefont {Zeng}},\
  }\bibfield  {title} {\bibinfo {title} {{Genuine tripartite nonlocality and
  entanglement in curved spacetime}},\ }\href
  {https://doi.org/10.1140/epjc/s10052-021-09954-4} {\bibfield  {journal}
  {\bibinfo  {journal} {Eur. Phys. J. C}\ }\textbf {\bibinfo {volume} {82}},\
  \bibinfo {pages} {4} (\bibinfo {year} {2022})},\ \Eprint
  {https://arxiv.org/abs/2201.02333} {arXiv:2201.02333 [quant-ph]} \BibitemShut
  {NoStop}%
\bibitem [{\citenamefont {Vidal}\ and\ \citenamefont
  {Werner}(2002)}]{Vidal:2002zz}%
  \BibitemOpen
  \bibfield  {author} {\bibinfo {author} {\bibfnamefont {G.}~\bibnamefont
  {Vidal}}\ and\ \bibinfo {author} {\bibfnamefont {R.~F.}\ \bibnamefont
  {Werner}},\ }\bibfield  {title} {\bibinfo {title} {{Computable measure of
  entanglement}},\ }\href {https://doi.org/10.1103/PhysRevA.65.032314}
  {\bibfield  {journal} {\bibinfo  {journal} {Phys. Rev. A}\ }\textbf {\bibinfo
  {volume} {65}},\ \bibinfo {pages} {032314} (\bibinfo {year} {2002})},\
  \Eprint {https://arxiv.org/abs/0102117} {arXiv:0102117 [quant-ph]}
  \BibitemShut {NoStop}%
\bibitem [{\citenamefont {D\'\i{}az}\ \emph {et~al.}(2023)\citenamefont
  {D\'\i{}az}, \citenamefont {Gonz\'alez}, \citenamefont {Hern\'andez},\ and\
  \citenamefont {Vergara}}]{Diaz:2023jrf}%
  \BibitemOpen
  \bibfield  {author} {\bibinfo {author} {\bibfnamefont {B.}~\bibnamefont
  {D\'\i{}az}}, \bibinfo {author} {\bibfnamefont {D.}~\bibnamefont
  {Gonz\'alez}}, \bibinfo {author} {\bibfnamefont {M.~J.}\ \bibnamefont
  {Hern\'andez}},\ and\ \bibinfo {author} {\bibfnamefont {J.~D.}\ \bibnamefont
  {Vergara}},\ }\bibfield  {title} {\bibinfo {title} {{Classical analogs of
  generalized purities, entropies, and logarithmic negativity}},\ }\href
  {https://doi.org/10.1103/PhysRevA.108.012411} {\bibfield  {journal} {\bibinfo
   {journal} {Phys. Rev. A}\ }\textbf {\bibinfo {volume} {108}},\ \bibinfo
  {pages} {012411} (\bibinfo {year} {2023})},\ \Eprint
  {https://arxiv.org/abs/2305.02887} {arXiv:2305.02887 [quant-ph]} \BibitemShut
  {NoStop}%
\bibitem [{\citenamefont {Xu}\ \emph {et~al.}(2024)\citenamefont {Xu},
  \citenamefont {Qi},\ and\ \citenamefont {Hou}}]{Xu:2024eqg}%
  \BibitemOpen
  \bibfield  {author} {\bibinfo {author} {\bibfnamefont {B.}~\bibnamefont
  {Xu}}, \bibinfo {author} {\bibfnamefont {X.}~\bibnamefont {Qi}},\ and\
  \bibinfo {author} {\bibfnamefont {J.}~\bibnamefont {Hou}},\ }\bibfield
  {title} {\bibinfo {title} {{Phase entanglement negativity for bipartite
  fermionic systems}},\ }\href {https://doi.org/10.1103/PhysRevA.110.032417}
  {\bibfield  {journal} {\bibinfo  {journal} {Phys. Rev. A}\ }\textbf {\bibinfo
  {volume} {110}},\ \bibinfo {pages} {032417} (\bibinfo {year}
  {2024})}\BibitemShut {NoStop}%
\bibitem [{\citenamefont {Wang}\ \emph
  {et~al.}(2010{\natexlab{b}})\citenamefont {Wang}, \citenamefont {Pan},\ and\
  \citenamefont {Jing}}]{Wang:2010bf}%
  \BibitemOpen
  \bibfield  {author} {\bibinfo {author} {\bibfnamefont {J.}~\bibnamefont
  {Wang}}, \bibinfo {author} {\bibfnamefont {Q.}~\bibnamefont {Pan}},\ and\
  \bibinfo {author} {\bibfnamefont {J.}~\bibnamefont {Jing}},\ }\bibfield
  {title} {\bibinfo {title} {{Entanglement redistribution in the Schwarzschild
  spacetime}},\ }\href {https://doi.org/10.1016/j.physletb.2010.07.035}
  {\bibfield  {journal} {\bibinfo  {journal} {Phys. Lett. B}\ }\textbf
  {\bibinfo {volume} {692}},\ \bibinfo {pages} {202} (\bibinfo {year}
  {2010}{\natexlab{b}})},\ \Eprint {https://arxiv.org/abs/1007.3331}
  {arXiv:1007.3331 [quant-ph]} \BibitemShut {NoStop}%
\bibitem [{\citenamefont {Ahn}\ \emph {et~al.}(2008)\citenamefont {Ahn},
  \citenamefont {Moon}, \citenamefont {Mann},\ and\ \citenamefont
  {Fuentes-Schuller}}]{Ahn:2008zf}%
  \BibitemOpen
  \bibfield  {author} {\bibinfo {author} {\bibfnamefont {D.}~\bibnamefont
  {Ahn}}, \bibinfo {author} {\bibfnamefont {Y.~H.}\ \bibnamefont {Moon}},
  \bibinfo {author} {\bibfnamefont {R.~B.}\ \bibnamefont {Mann}},\ and\
  \bibinfo {author} {\bibfnamefont {I.}~\bibnamefont {Fuentes-Schuller}},\
  }\bibfield  {title} {\bibinfo {title} {{The Black hole final state for the
  Dirac fields In Schwarzschild spacetime}},\ }\href
  {https://doi.org/10.1088/1126-6708/2008/06/062} {\bibfield  {journal}
  {\bibinfo  {journal} {JHEP}\ }\textbf {\bibinfo {volume} {06}},\ \bibinfo
  {pages} {062}},\ \Eprint {https://arxiv.org/abs/0801.0471} {arXiv:0801.0471
  [hep-th]} \BibitemShut {NoStop}%
\bibitem [{\citenamefont {Martin-Martinez}\ and\ \citenamefont
  {Leon}(2010)}]{Martin-Martinez:2010bcj}%
  \BibitemOpen
  \bibfield  {author} {\bibinfo {author} {\bibfnamefont {E.}~\bibnamefont
  {Martin-Martinez}}\ and\ \bibinfo {author} {\bibfnamefont {J.}~\bibnamefont
  {Leon}},\ }\bibfield  {title} {\bibinfo {title} {{Quantum correlations
  through event horizons: Fermionic versus bosonic entanglement}},\ }\href
  {https://doi.org/10.1103/PhysRevA.81.032320} {\bibfield  {journal} {\bibinfo
  {journal} {Phys. Rev. A}\ }\textbf {\bibinfo {volume} {81}},\ \bibinfo
  {pages} {032320} (\bibinfo {year} {2010})},\ \Eprint
  {https://arxiv.org/abs/1001.4302} {arXiv:1001.4302 [quant-ph]} \BibitemShut
  {NoStop}%
\bibitem [{\citenamefont {Kalb}\ and\ \citenamefont
  {Ramond}(1974)}]{Kalb:1974yc}%
  \BibitemOpen
  \bibfield  {author} {\bibinfo {author} {\bibfnamefont {M.}~\bibnamefont
  {Kalb}}\ and\ \bibinfo {author} {\bibfnamefont {P.}~\bibnamefont {Ramond}},\
  }\bibfield  {title} {\bibinfo {title} {{Classical direct interstring
  action}},\ }\href {https://doi.org/10.1103/PhysRevD.9.2273} {\bibfield
  {journal} {\bibinfo  {journal} {Phys. Rev. D}\ }\textbf {\bibinfo {volume}
  {9}},\ \bibinfo {pages} {2273} (\bibinfo {year} {1974})}\BibitemShut
  {NoStop}%
\bibitem [{\citenamefont {Yang}\ \emph {et~al.}(2023)\citenamefont {Yang},
  \citenamefont {Chen}, \citenamefont {Duan},\ and\ \citenamefont
  {Zhao}}]{Yang:2023wtu}%
  \BibitemOpen
  \bibfield  {author} {\bibinfo {author} {\bibfnamefont {K.}~\bibnamefont
  {Yang}}, \bibinfo {author} {\bibfnamefont {Y.-Z.}\ \bibnamefont {Chen}},
  \bibinfo {author} {\bibfnamefont {Z.-Q.}\ \bibnamefont {Duan}},\ and\
  \bibinfo {author} {\bibfnamefont {J.-Y.}\ \bibnamefont {Zhao}},\ }\bibfield
  {title} {\bibinfo {title} {{Static and spherically symmetric black holes in
  gravity with a background Kalb-Ramond field}},\ }\href
  {https://doi.org/10.1103/PhysRevD.108.124004} {\bibfield  {journal} {\bibinfo
   {journal} {Phys. Rev. D}\ }\textbf {\bibinfo {volume} {108}},\ \bibinfo
  {pages} {124004} (\bibinfo {year} {2023})},\ \Eprint
  {https://arxiv.org/abs/2308.06613} {arXiv:2308.06613 [gr-qc]} \BibitemShut
  {NoStop}%
\bibitem [{\citenamefont {Duan}\ \emph {et~al.}(2024)\citenamefont {Duan},
  \citenamefont {Zhao},\ and\ \citenamefont {Yang}}]{Duan:2023gng}%
  \BibitemOpen
  \bibfield  {author} {\bibinfo {author} {\bibfnamefont {Z.-Q.}\ \bibnamefont
  {Duan}}, \bibinfo {author} {\bibfnamefont {J.-Y.}\ \bibnamefont {Zhao}},\
  and\ \bibinfo {author} {\bibfnamefont {K.}~\bibnamefont {Yang}},\ }\bibfield
  {title} {\bibinfo {title} {{Electrically charged black holes in gravity with
  a background Kalb\textendash{}Ramond field}},\ }\href
  {https://doi.org/10.1140/epjc/s10052-024-13188-5} {\bibfield  {journal}
  {\bibinfo  {journal} {Eur. Phys. J. C}\ }\textbf {\bibinfo {volume} {84}},\
  \bibinfo {pages} {798} (\bibinfo {year} {2024})},\ \Eprint
  {https://arxiv.org/abs/2310.13555} {arXiv:2310.13555 [gr-qc]} \BibitemShut
  {NoStop}%
\bibitem [{\citenamefont {Liu}\ \emph {et~al.}(2024{\natexlab{c}})\citenamefont
  {Liu}, \citenamefont {Wu},\ and\ \citenamefont {Wang}}]{Liu:2024oas}%
  \BibitemOpen
  \bibfield  {author} {\bibinfo {author} {\bibfnamefont {W.}~\bibnamefont
  {Liu}}, \bibinfo {author} {\bibfnamefont {D.}~\bibnamefont {Wu}},\ and\
  \bibinfo {author} {\bibfnamefont {J.}~\bibnamefont {Wang}},\ }\bibfield
  {title} {\bibinfo {title} {{Static neutral black holes in Kalb-Ramond
  gravity}},\ }\href {https://doi.org/10.1088/1475-7516/2024/09/017} {\bibfield
   {journal} {\bibinfo  {journal} {JCAP}\ }\textbf {\bibinfo {volume} {09}},\
  \bibinfo {pages} {017}},\ \Eprint {https://arxiv.org/abs/2406.13461}
  {arXiv:2406.13461 [hep-th]} \BibitemShut {NoStop}%
\bibitem [{\citenamefont {Liu}\ \emph {et~al.}(2024{\natexlab{d}})\citenamefont
  {Liu}, \citenamefont {Wu},\ and\ \citenamefont {Wang}}]{Liu:2024lve}%
  \BibitemOpen
  \bibfield  {author} {\bibinfo {author} {\bibfnamefont {W.}~\bibnamefont
  {Liu}}, \bibinfo {author} {\bibfnamefont {D.}~\bibnamefont {Wu}},\ and\
  \bibinfo {author} {\bibfnamefont {J.}~\bibnamefont {Wang}},\ }\bibfield
  {title} {\bibinfo {title} {{Shadow of slowly rotating Kalb-Ramond black
  holes}},\ }\href@noop {} {\  (\bibinfo {year} {2024}{\natexlab{d}})},\
  \Eprint {https://arxiv.org/abs/2407.07416} {arXiv:2407.07416 [gr-qc]}
  \BibitemShut {NoStop}%
\bibitem [{\citenamefont {Henderson}\ \emph {et~al.}(2018)\citenamefont
  {Henderson}, \citenamefont {Hennigar}, \citenamefont {Mann}, \citenamefont
  {Smith},\ and\ \citenamefont {Zhang}}]{Henderson:2017yuv}%
  \BibitemOpen
  \bibfield  {author} {\bibinfo {author} {\bibfnamefont {L.~J.}\ \bibnamefont
  {Henderson}}, \bibinfo {author} {\bibfnamefont {R.~A.}\ \bibnamefont
  {Hennigar}}, \bibinfo {author} {\bibfnamefont {R.~B.}\ \bibnamefont {Mann}},
  \bibinfo {author} {\bibfnamefont {A.~R.~H.}\ \bibnamefont {Smith}},\ and\
  \bibinfo {author} {\bibfnamefont {J.}~\bibnamefont {Zhang}},\ }\bibfield
  {title} {\bibinfo {title} {{Harvesting Entanglement from the Black Hole
  Vacuum}},\ }\href {https://doi.org/10.1088/1361-6382/aae27e} {\bibfield
  {journal} {\bibinfo  {journal} {Class. Quant. Grav.}\ }\textbf {\bibinfo
  {volume} {35}},\ \bibinfo {pages} {21LT02} (\bibinfo {year} {2018})},\
  \Eprint {https://arxiv.org/abs/1712.10018} {arXiv:1712.10018 [quant-ph]}
  \BibitemShut {NoStop}%
\bibitem [{\citenamefont {Cong}\ \emph {et~al.}(2020)\citenamefont {Cong},
  \citenamefont {Qian}, \citenamefont {Good},\ and\ \citenamefont
  {Mann}}]{Cong:2020nec}%
  \BibitemOpen
  \bibfield  {author} {\bibinfo {author} {\bibfnamefont {W.}~\bibnamefont
  {Cong}}, \bibinfo {author} {\bibfnamefont {C.}~\bibnamefont {Qian}}, \bibinfo
  {author} {\bibfnamefont {M.~R.~R.}\ \bibnamefont {Good}},\ and\ \bibinfo
  {author} {\bibfnamefont {R.~B.}\ \bibnamefont {Mann}},\ }\bibfield  {title}
  {\bibinfo {title} {{Effects of Horizons on Entanglement Harvesting}},\ }\href
  {https://doi.org/10.1007/JHEP10(2020)067} {\bibfield  {journal} {\bibinfo
  {journal} {JHEP}\ }\textbf {\bibinfo {volume} {10}},\ \bibinfo {pages}
  {067}},\ \Eprint {https://arxiv.org/abs/2006.01720} {arXiv:2006.01720
  [gr-qc]} \BibitemShut {NoStop}%
\bibitem [{\citenamefont {Liu}\ \emph {et~al.}(2023{\natexlab{b}})\citenamefont
  {Liu}, \citenamefont {Zhang},\ and\ \citenamefont {Yu}}]{Liu:2022uhf}%
  \BibitemOpen
  \bibfield  {author} {\bibinfo {author} {\bibfnamefont {Z.}~\bibnamefont
  {Liu}}, \bibinfo {author} {\bibfnamefont {J.}~\bibnamefont {Zhang}},\ and\
  \bibinfo {author} {\bibfnamefont {H.}~\bibnamefont {Yu}},\ }\bibfield
  {title} {\bibinfo {title} {{Entanglement harvesting of accelerated detectors
  versus static ones in a thermal bath}},\ }\href
  {https://doi.org/10.1103/PhysRevD.107.045010} {\bibfield  {journal} {\bibinfo
   {journal} {Phys. Rev. D}\ }\textbf {\bibinfo {volume} {107}},\ \bibinfo
  {pages} {045010} (\bibinfo {year} {2023}{\natexlab{b}})},\ \Eprint
  {https://arxiv.org/abs/2208.14825} {arXiv:2208.14825 [quant-ph]} \BibitemShut
  {NoStop}%
\bibitem [{\citenamefont {Maeso-Garc\'\i{}a}\ \emph {et~al.}(2023)\citenamefont
  {Maeso-Garc\'\i{}a}, \citenamefont {Polo-G\'omez},\ and\ \citenamefont
  {Mart\'\i{}n-Mart\'\i{}nez}}]{Maeso-Garcia:2022uzf}%
  \BibitemOpen
  \bibfield  {author} {\bibinfo {author} {\bibfnamefont {H.}~\bibnamefont
  {Maeso-Garc\'\i{}a}}, \bibinfo {author} {\bibfnamefont {J.}~\bibnamefont
  {Polo-G\'omez}},\ and\ \bibinfo {author} {\bibfnamefont {E.}~\bibnamefont
  {Mart\'\i{}n-Mart\'\i{}nez}},\ }\bibfield  {title} {\bibinfo {title} {{How
  measuring a quantum field affects entanglement harvesting}},\ }\href
  {https://doi.org/10.1103/PhysRevD.107.045011} {\bibfield  {journal} {\bibinfo
   {journal} {Phys. Rev. D}\ }\textbf {\bibinfo {volume} {107}},\ \bibinfo
  {pages} {045011} (\bibinfo {year} {2023})},\ \Eprint
  {https://arxiv.org/abs/2210.05692} {arXiv:2210.05692 [quant-ph]} \BibitemShut
  {NoStop}%
\bibitem [{\citenamefont {Liu}\ \emph {et~al.}(2023{\natexlab{c}})\citenamefont
  {Liu}, \citenamefont {Wu}, \citenamefont {Wen},\ and\ \citenamefont
  {Wang}}]{Liu:2023awu}%
  \BibitemOpen
  \bibfield  {author} {\bibinfo {author} {\bibfnamefont {Q.}~\bibnamefont
  {Liu}}, \bibinfo {author} {\bibfnamefont {S.-M.}\ \bibnamefont {Wu}},
  \bibinfo {author} {\bibfnamefont {C.}~\bibnamefont {Wen}},\ and\ \bibinfo
  {author} {\bibfnamefont {J.}~\bibnamefont {Wang}},\ }\bibfield  {title}
  {\bibinfo {title} {{Quantum properties of fermionic fields in multi-event
  horizon spacetime}},\ }\href {https://doi.org/10.1007/s11433-023-2246-8}
  {\bibfield  {journal} {\bibinfo  {journal} {Sci. China Phys. Mech. Astron.}\
  }\textbf {\bibinfo {volume} {66}},\ \bibinfo {pages} {120413} (\bibinfo
  {year} {2023}{\natexlab{c}})},\ \Eprint {https://arxiv.org/abs/2311.07047}
  {arXiv:2311.07047 [gr-qc]} \BibitemShut {NoStop}%
\bibitem [{\citenamefont {Liu}\ \emph {et~al.}(2023{\natexlab{d}})\citenamefont
  {Liu}, \citenamefont {Zhang},\ and\ \citenamefont {Yu}}]{Liu:2023zro}%
  \BibitemOpen
  \bibfield  {author} {\bibinfo {author} {\bibfnamefont {Z.}~\bibnamefont
  {Liu}}, \bibinfo {author} {\bibfnamefont {J.}~\bibnamefont {Zhang}},\ and\
  \bibinfo {author} {\bibfnamefont {H.}~\bibnamefont {Yu}},\ }\bibfield
  {title} {\bibinfo {title} {{Harvesting correlations from vacuum quantum
  fields in the presence of a reflecting boundary}},\ }\href
  {https://doi.org/10.1007/JHEP11(2023)184} {\bibfield  {journal} {\bibinfo
  {journal} {JHEP}\ }\textbf {\bibinfo {volume} {11}},\ \bibinfo {pages}
  {184}},\ \Eprint {https://arxiv.org/abs/2310.07164} {arXiv:2310.07164
  [quant-ph]} \BibitemShut {NoStop}%
\bibitem [{\citenamefont {Lindel}\ \emph {et~al.}(2024)\citenamefont {Lindel},
  \citenamefont {Herter}, \citenamefont {Gebhart}, \citenamefont {Faist},\ and\
  \citenamefont {Buhmann}}]{Lindel:2023rfi}%
  \BibitemOpen
  \bibfield  {author} {\bibinfo {author} {\bibfnamefont {F.}~\bibnamefont
  {Lindel}}, \bibinfo {author} {\bibfnamefont {A.}~\bibnamefont {Herter}},
  \bibinfo {author} {\bibfnamefont {V.}~\bibnamefont {Gebhart}}, \bibinfo
  {author} {\bibfnamefont {J.}~\bibnamefont {Faist}},\ and\ \bibinfo {author}
  {\bibfnamefont {S.~Y.}\ \bibnamefont {Buhmann}},\ }\bibfield  {title}
  {\bibinfo {title} {{Entanglement harvesting from electromagnetic quantum
  fields}},\ }\href {https://doi.org/10.1103/PhysRevA.110.022414} {\bibfield
  {journal} {\bibinfo  {journal} {Phys. Rev. A}\ }\textbf {\bibinfo {volume}
  {110}},\ \bibinfo {pages} {022414} (\bibinfo {year} {2024})},\ \Eprint
  {https://arxiv.org/abs/2311.04642} {arXiv:2311.04642 [quant-ph]} \BibitemShut
  {NoStop}%
\bibitem [{\citenamefont {Ji}\ \emph {et~al.}(2024)\citenamefont {Ji},
  \citenamefont {Zhang},\ and\ \citenamefont {Yu}}]{Ji:2024fcq}%
  \BibitemOpen
  \bibfield  {author} {\bibinfo {author} {\bibfnamefont {Y.}~\bibnamefont
  {Ji}}, \bibinfo {author} {\bibfnamefont {J.}~\bibnamefont {Zhang}},\ and\
  \bibinfo {author} {\bibfnamefont {H.}~\bibnamefont {Yu}},\ }\bibfield
  {title} {\bibinfo {title} {{Entanglement harvesting in cosmic string
  spacetime}},\ }\href {https://doi.org/10.1007/JHEP06(2024)161} {\bibfield
  {journal} {\bibinfo  {journal} {JHEP}\ }\textbf {\bibinfo {volume} {06}},\
  \bibinfo {pages} {161}},\ \Eprint {https://arxiv.org/abs/2401.13406}
  {arXiv:2401.13406 [quant-ph]} \BibitemShut {NoStop}%
\bibitem [{\citenamefont {Wu}\ \emph {et~al.}(2024{\natexlab{c}})\citenamefont
  {Wu}, \citenamefont {Wang}, \citenamefont {Huang},\ and\ \citenamefont
  {Wang}}]{Wu:2024whx}%
  \BibitemOpen
  \bibfield  {author} {\bibinfo {author} {\bibfnamefont {S.-M.}\ \bibnamefont
  {Wu}}, \bibinfo {author} {\bibfnamefont {R.-D.}\ \bibnamefont {Wang}},
  \bibinfo {author} {\bibfnamefont {X.-L.}\ \bibnamefont {Huang}},\ and\
  \bibinfo {author} {\bibfnamefont {Z.}~\bibnamefont {Wang}},\ }\bibfield
  {title} {\bibinfo {title} {{Harvesting asymmetric steering via non-identical
  detectors}},\ }\href@noop {} {\  (\bibinfo {year} {2024}{\natexlab{c}})},\
  \Eprint {https://arxiv.org/abs/2408.11277} {arXiv:2408.11277 [quant-ph]}
  \BibitemShut {NoStop}%
\bibitem [{\citenamefont {Zeng}\ \emph {et~al.}(2024)\citenamefont {Zeng},
  \citenamefont {Liu},\ and\ \citenamefont {Wu}}]{Zeng:2024qme}%
  \BibitemOpen
  \bibfield  {author} {\bibinfo {author} {\bibfnamefont {H.-S.}\ \bibnamefont
  {Zeng}}, \bibinfo {author} {\bibfnamefont {H.}~\bibnamefont {Liu}},\ and\
  \bibinfo {author} {\bibfnamefont {L.-J.}\ \bibnamefont {Wu}},\ }\bibfield
  {title} {\bibinfo {title} {{Schwinger correlation of Dirac fields in
  accelerated frames}},\ }\href {https://doi.org/10.1088/1361-6382/ad3ac8}
  {\bibfield  {journal} {\bibinfo  {journal} {Class. Quant. Grav.}\ }\textbf
  {\bibinfo {volume} {41}},\ \bibinfo {pages} {115006} (\bibinfo {year}
  {2024})}\BibitemShut {NoStop}%
\end{thebibliography}

\end{document}